\documentclass{elsart}

\usepackage{amssymb}
\usepackage[dvips]{graphicx}
\usepackage{dcolumn}
\usepackage{bm}
\usepackage{amsmath}

\begin{document}

\begin{frontmatter}


\title{Dual vortex theory of doped Mott insulators}

\author[UCSB]{Leon Balents} and
\author[Harvard]{Subir Sachdev}

\address[UCSB]{Department of Physics, University of California, Santa
Barbara, CA 93106-4030}

\address[Harvard]{Department of Physics, Harvard University, Cambridge MA
02138}

\begin{abstract}
We present a general framework for describing the quantum phases
obtained by doping paramagnetic Mott insulators on the square
lattice. The undoped insulators are efficiently characterized by the
projective transformations of various fields under the square
lattice space group (the PSG). We show that the PSG also imposes
powerful constraints on the doped system, and on the effective
action for the vortex and Bogoliubov quasiparticle excitations of
superconducting states. This action can also be extended across
transitions to supersolid or insulating states at nonzero doping.
For the case of a valence bond solid (VBS) insulator, we show that
the doped system has the same PSG as that of elementary bosons with
density equal to the density of electron Cooper pairs. We also
discuss aspects of the action for a $d$-wave superconductor obtained
by doping a ``staggered-flux'' spin liquid state.
\end{abstract}


\end{frontmatter}

\section{Introduction}
\label{sec:intro}

It is now widely accepted that the cuprate superconductors can be
understood as doped Mott insulators. Anderson \cite{pwa} made the early
suggestion that appropriate reference Mott insulating ground state
should be paramagnetic {\em i.e.} preserve spin rotation invariance with
the electron spin operators obeying $\langle {\bf S}_j \rangle = 0$ on
all sites, $j$, of the square lattice; these are also loosely referred
to as the resonating valence bond (RVB) states of Pauling
\cite{pauling}, or as spin liquids. A more subtle and complete
understanding of spin liquids on the square lattice has emerged since
then. It has been found that some of the physically interesting spin
liquids are unstable at low energies to confinement and the emergence of
new competing order parameters.  The best known examples are $U(1)$ RVB
states with a gap to both spin and charge excitations, which are
unstable by general arguments.  The instability of the simplest such
state leads to the prominent example of competing valence bond solid
(VBS) order\cite{rs}, which will play an important role in
the considerations of this paper. Other spin liquids can potentially be
stable against symmetry breaking, \cite{hermele,asl} but instead have
gapless gauge and fermionic `spinon' excitations which are strongly
coupled together in an `algebraic spin liquid': we will consider the
example of the `staggered flux' (sF) state in this paper. The sF state
also has strongly enhanced fluctuations of the VBS and other orders.

Many authors have described how doping a paramagnetic Mott
insulator with charged carriers leads to $d$-wave
superconductivity. \cite{kotliar,fukuyama,sr,su2,vs} This leads to a
number of natural questions on the role of competing orders, such as
VBS or charge density wave orders, in the doped system. ({\em i\/})
Can the competing orders have long-range order in the doped system,
and how does the order parameter at finite doping relate to the VBS
order in the undoped insulator~? ({\em ii\/}) Does the long-range
order co-exist with superconductivity, or can it appear only in
finite doping insulating states~? ({\em iii\/}) What are the
theories of the quantum critical points between such states, and can
they be of the Landau-forbidden\cite{senthil1,senthil2,senthil3}
variety~? ({\em iv\/}) In the moderate doping $d$-wave
superconductor, where this is no competing long-range order, what
are the correlations of the competing order, and what is their
interplay with vortices and quasiparticles~? This paper will develop
a formalism designed to address such questions. We will do this in
as general a setting as possible, relying mainly on symmetry
arguments. We will not use any specific microscopic model of the
doped antiferromagnet, and so will not make any quantitative
predictions of the phase diagram.

Our analysis will focus on two distinct classes of paramagnetic,
undoped Mott insulators. In both classes, the Mott insulator is
described by the dynamics of a compact U(1) gauge field
$\mathcal{A}_{\mu}$ ($\mu = x,y,\tau$ is a spacetime index). The
``photon'' excitation of this gauge field describes $S=0$
excitations above a spin singlet RVB ground state. Over at least a
significant intermediate energy range, this gauge field can be
described by a conventional, Maxwell action of the form
\begin{equation}
\mathcal{S}_{\mathcal{A}} = - K \sum_{\Box} \cos \left(
\epsilon_{\mu\nu\lambda} \Delta_\nu \mathcal{A}_{j \lambda} \right).
\label{sa}
\end{equation}
We have discretized spacetime onto the sites of a cubic lattice,
$j$, with each equal time section mapping onto the sites of the
underlying physical square lattice (it should be clear from the
context whether $j$ refers to the square or cubic lattice), $\Box$
represents a sum over elementary plaquettes of the cubic lattice,
$\Delta_\mu$ is a lattice derivative, the indices $\mu$, $\nu$,
$\lambda$ extend over $x$, $y$, $\tau$, and $K$ is a coupling
constant. In addition to the pure gauge fluctuations in
Eq.~(\ref{sa}), the theory of the undoped Mott insulator must also
account for the `matter' fields: the matter associated with the
average density of exactly one electron per site. For our purposes,
it turns out that the most important effect \cite{haldane,rs} of the
matter in the undoped insulator is on its consequences for the {\em
monopoles\/} in $\mathcal{A}_\mu$ (the monopoles are discrete
tunnelling events in which the net U(1) gauge flux is changed by $2
\pi$). The matter causes each monopole to transform non-trivially
under the space group symmetries of the square lattice, and we dub
these transformations (following Wen, \cite{wen} in a slight abuse
of mathematical terminology) as the PSG (for projective symmetry
group) of the monopoles.\cite{senthil2,psgbosons,psgdimers} The
monopole PSG specifies the competing order parameters which emerge
from the RVB state at low energies.

We turn next to the doped antiferromagnet, where the ground state can be
a superconductor. Here, one class of excitations of the superconductor
are the vortices in Cooper pair condensate, each carrying
electromagnetic flux of $hc/(2 e)$. It was shown in recent work
\cite{psgbosons,psgdimers} that a general description of competing
orders near to a superconducting state is obtained by analyzing the PSG
of the vortices. The PSG ties each vortex to fluctuations in competing
orders associated with generalized `density-wave' and `vorticity-wave'
modulations, and a number of observable consequences of this connection
were discussed.

So how do we describe the evolution of the competing orders in the
undoped Mott insulator to those in the superconductor~? We will
argue in this paper that a very powerful and general method of doing
this is provided by the PSG. Quite simply, we combine the PSG of the
monopoles of the undoped Mott insulator with the PSG of the vortices
in the doped system, and obtain a combined action which is invariant
under both PSGs. In principle, this effective action applies in both
the undoped insulator and the highly doped superconductors, and at
all doping concentrations in between. It is designed to address the
questions posed in the second paragraph.

$ $From a different perspective, the approach of this paper is an
expansion about a ``Mott quantum critical point'', describing a
system on the verge of superconductivity due to charge condensation
in a non-trivial paramagnetic Mott insulator.  The crucial
assumption is the validity of this expansion, i.e the closeness in
phase space of the physical system to the quantum critical point.
The degree of stability of the non-trivial Mott insulator -- the
``reference state'' -- is a secondary issue.  Indeed, in the first
case we consider, the non-trivial Mott insulator is the $U(1)$ RVB
state with fully gapped spin and charge excitations, which is
generically unstable to VBS order at low energies. We therefore
denote this RVB state as a VBS insulator, though most of our
considerations never require the actual occurence of long-range VBS
order of any particular type.  Nevertheless, a number of recent
works have shown that such a state can be a good starting point for
describing the quantum critical regime in which VBS order is weak.
The different possible paramagnetic Mott states do, however,
generally lead to physically different predictions.  A proper choice
amongst such states must be made on energetic or phenomenological
(empirical) grounds.

The following two subsections briefly introduce the two classes of
paramagnetic Mott insulators we consider. For the VBS insulator,
introduced in Section~\ref{vbs1}, we will be able to carry our
derivation of the action to completion, using the PSG of the
monopoles described in Ref.~\cite{senthil2}. The sF state will be
considered in Section~\ref{sf1}: the monopole PSG of this state has
not been yet been computed and so our analysis of this case will
remain incomplete. Nevertheless, we will be able to make substantial
progress in describing the structure of the action in this paper,
and note significant differences from the VBS case. The final
subsection of this introduction will briefly outline the strategy
for extending our analysis to finite doping, and discuss the issue
of flux quantization.

\subsection{The Valence Bond Solid Insulator}
\label{vbs1}

Our first quantum paramagnet is obtained in a theory of quantum
fluctuations about an antiferromagnetically ordered N\'{e}el state.
This theory focusses on the N\'{e}el order parameter as the primary
dynamic degree of freedom, and obtains VBS order in the state where
quantum fluctuations have ``disordered'' the N\'{e}el order.
\cite{rs,paris}

The N\'eel state is described by the N\'eel order parameter, ${\bf
n}_j$, which is a unit vector in spin space, related to the electron
spin operator by
\begin{equation}
{\bf S}_j = S \eta_j {\bf n}_j, \label{defneel}
\end{equation}
where $S=1/2$ is the spin, and $\eta_j = \pm 1$ takes opposite signs
on the two sublattices. The action for quantum spin fluctuations of
the ${\bf n}$ has a familiar description in terms of the O(3)
non-linear $\sigma$ model with Berry phases:
\begin{equation}
\mathcal{S}_{\bf n} = \frac{1}{2g} \sum_j \left( \Delta_\mu {\bf n}
\right)^2 + i S \sum_j \eta_j  {\bf A}_D ({\bf n}_j ) \cdot
\Delta_\tau {\bf n}_j , \label{nls1}
\end{equation}
where ${\bf A}_D ( {\bf n}) = {\bf A}_D (- {\bf n})$ is the Dirac
monopole function which is used to yield a Berry phase proportional
to the area enclosed by the worldline of each spin. The coupling
constant $g$ can be tune to `disorder' the N\'eel state: for small
$g$, the ground state has N\'eel order with $\langle {\bf n} \rangle
\neq 0$, while for large $g$, there are strong fluctuations of ${\bf
n}$ leading to a RVB state with $\langle {\bf n} \rangle =0$.
However, this ``disordered'' state nevertheless has non-trivial
quantum dynamics which is expressed most easily in terms of the
gauge field $\mathcal{A}_\mu$ which is defined here by
\begin{equation}
\mathcal{A}_\mu \equiv \frac{1}{2} {\bf A}_D ({\bf n}) \cdot
\Delta_\mu {\bf n}. \label{nls2}
\end{equation}
As has been argued elsewhere, \cite{paris} for large $g$, the
effective action for $\mathcal{A}_\mu$ in this RVB state is
$\mathcal{S}_{\mathcal{A}} + \mathcal{S}_B$, where
$\mathcal{S}_{\mathcal{A}}$ was in Eq.~(\ref{sa}) and the Berry
phase term
\begin{equation}
\mathcal{S}_B = i 2S \sum_j \mathcal{A}_{j\tau} \label{berry1}
\end{equation}
descends directly from the second term in Eq.~(\ref{nls1}). The
properties of $\mathcal{S}_{\mathcal{A}} + \mathcal{S}_B$ have been
described in detail elsewhere,\cite{rs,sj} and will be reviewed here
in Section~\ref{qdn1}. Briefly, the Berry phase endow the monopoles
in $\mathcal{A}_\mu$ with non-trivial transformations under the PSG,
so that they transform like a VBS order parameter.\cite{rs,senthil2}
Further, at low energies, the monopoles ``condense'' leading to VBS
order in the ground state. The full details of the PSG will be
presented in Section~\ref{qdn}, but here we note the transformation
of $\mathcal{A}_\mu$ under the time-reversal operation,
$\mathcal{T}$. This can be deduced from Eqs.~(\ref{nls1}) and
(\ref{nls2}): we have ${\bf n} \rightarrow - {\bf n}$, and hence
\begin{equation}
\mathcal{T}~:~\mathcal{A}_\tau \rightarrow
\mathcal{A}_\tau~;~\mathcal{A}_x \rightarrow
-\mathcal{A}_x~;~\mathcal{A}_y \rightarrow -\mathcal{A}_y.
\label{ta1}
\end{equation}
We will observe shortly that $\mathcal{A}_\mu$ has a distinct
transformation under $\mathcal{T}$ for our second quantum
paramagnet: this distinctions plays an important role in the two
theories of vortex dynamics in the $d$-wave superconductor.

\subsection{The Staggered Flux spin liquid} \label{sf1}

This RVB state, which we denote by the sF spin liquid, has seen much
recent discussion in the literature,\cite{hermele,asl,su2,nagaosa} and
its basic properties will be reviewed in Section~\ref{sf2}.  Here, we
note that the low energy (fixed point) properties of the sF spin liquid
are described by $\mathcal{S}_\mathcal{A} + \mathcal{S}_\Psi$, where
$\mathcal{S}_\Psi$ describes the coupling of $\mathcal{A}_\mu$ to 4
two-component, fermionic, massless Dirac fields, $\Psi$; schematically,
the action is $\mathcal{S}_\Psi = \int d^2 r d\tau \mathcal{L}_\Psi
[\mathcal{A}_\mu]$, where
\begin{equation}
\mathcal{L}_\Psi [ \mathcal{A}_\mu]  =  -i \overline{\Psi}
\gamma_\mu \left(
\partial_\mu + i\mathcal{A}_\mu \right) \Psi, \label{sdirac}
\end{equation}
where $\gamma_\mu$ are the Dirac matrices. The PSG properties of the
sF spin liquid are described in Section~\ref{sf3}. Under
$\mathcal{T}$, the gauge field $\mathcal{A}_\mu$ transforms, as
expected from the relativistic structure of $\mathcal{S}_\Psi$, as a
conventional Lorentz vector
\begin{equation}
\mathcal{T}~:~\mathcal{A}_\tau \rightarrow
-\mathcal{A}_\tau~;~\mathcal{A}_x \rightarrow
\mathcal{A}_x~;~\mathcal{A}_y \rightarrow \mathcal{A}_y. \label{ta2}
\end{equation}
Note the distinction of this from Eq.~(\ref{ta1}) for the VBS state.
Monopoles also play an important role in dynamics of the sF spin
liquid. \cite{ioffe,marston,wen,hermele} Computing their PSG remains
an important open problem, and so we will perform our analysis to
the extent possible, pending the eventual determination of the
monopole PSG.

\subsection{Doping the antiferromagnet}
\label{dta}

A detailed description of the theory obtained by doping the two
paramagnetic Mott insulators described above appears in
Sections~\ref{qdn} and~\ref{sf2} respectively. Here, we note some
important common features of the two theories.

In both cases, we represent the charge carriers by {\em two} species
\cite{su2,vortex,psgdimers} of spinless bosons\footnote{A
complementary theory with spinless fermionic holons, and bosonic
spinons, is discussed in Ref.~\cite{dcpdoped}. Such a theory is
natural when the undoped VBS state is proximate to a deconfined
critical point \cite{senthil1,senthil2,senthil3} to a magnetically
ordered N\'eel state. The approach in the present paper to doping
the VBS state with bosonic holons is more closely connected to the
large $N$ limit studied in Ref.~\cite{vs}.} (`holons'), $b_1$ and
$b_2$. If the total hole density is $\delta$, then each species of
boson has density $\delta/2$, and we will assume throughout that
\begin{equation}
\frac{\delta}{2} = \frac{p}{q} \label{defpq}
\end{equation}
where $p$ and $q$ are relatively prime integers. As we will see in
the body of the paper, the number-theoretic properties of the
integer $q$ will play a crucial role in our analysis.

Both $b_1$ and $b_2$ carry physical electromagnetic charge $e$.
However, they carry opposite charges $\pm 1$ under the internal
compact U(1) gauge field $\mathcal{A}_\mu$. The superconducting
state is obtained when the charge $2e$, gauge neutral, combination
$b_1 b_2$ condenses. However, with charge $e$ bosons present in the
theory, one may wonder if there are regions of stability of flux
$hc/e$ vortices. As we review below, quite generally all stable
vortices have flux $hc/(2e)$.

Far enough from the vortex cores, we may focus on the phases of the
boson fields alone. Let us therefore write $b_1 \sim e^{-i
\theta_1}$ and $b_2 \sim e^{-i \theta_2}$. We denote a vortex in
which the phase of $\theta_1$ winds by $2 \pi$ by $\psi_1$, and a
vortex in which the phase of $\theta_2$ winds by $2 \pi$ by
$\psi_2$. A complete dual theory of the $\psi_{1,2}$ vortices
appears in subsequent sections. Here, we continue with an analysis
in the direct picture. Far from the vortex core, the free energy
contains a contribution
\begin{equation}
F = \frac{\rho_s}{2} \int d^2 r \left[ \left(\vec{\nabla} \theta_1 -
\vec{\mathcal{A}} \right)^2 + \left(\vec{\nabla} \theta_2 +
\vec{\mathcal{A}} \right)^2 \right]. \label{Fv}
\end{equation}
We have not included the electromagnetic magnetic field here, as its
influence is important only at distances of order the London
penetration depth. For a $\psi_1$ vortex, $\vec{\nabla} \theta_1 =
\vec{e}_\theta /r$, and $\vec{\nabla} \theta_2 = 0$ at large $r$
($\vec{e}_\theta$ is a unit vector orthogonal to the radial
direction). The slow $1/r$ decay implies a logarithmically divergent
vortex energy in Eq.~(\ref{Fv}) which is cut off at the London
penetration depth. Let us assume that $\vec{\mathcal{A}} = \alpha
\vec{e}_\theta / r$ at large $r$ (which is pure gauge). Then the
co-efficient of the logarithmically divergent term is proportional
to $(1- \alpha)^2 + \alpha^2$. Minimizing w.r.t. to $\alpha$, we
obtain $\alpha = 1/2$. Thus a $\psi_1$ vortex carries a total
$\mathcal{A}_\mu$ gauge flux $=\oint \mathcal{A}_\mu  dr_\mu = \pi$.
Similarly, a $\psi_2$ vortex carries $\mathcal{A}_\mu$ gauge flux of
$-\pi$. In both cases, at distances of order the London penetration
depth, there must be a residual gauge flux $\pi$ (in units of $\hbar
c /e$) in the physical electromagnetic gauge field to ultimately
render the energy of a vortex finite. So both the $\psi_1$ and
$\psi_2$ vortices carry flux $hc/(2e)$.

The above $\mathcal{A}_\mu$ flux assignments also allow us to
identify the connection between the vortices and the monopoles. A
monopole is a tunneling event in which the $\mathcal{A}_\mu$ flux
changes by $2 \pi$. Such a change in flux also occurs when a
$\psi_1$ vortex transforms into a $\psi_2$ vortex. At low energies
we expect the two tunneling events to always co-incide: this is
especially so in the superconducting state, where the
$\mathcal{A}_\mu$ gauge flux is confined by the $b_{1,2}$
condensate. So the operator for a monopole tunneling event is
$\psi_2^\dagger \psi_1$.

The following Section~\ref{qdn} will present our theory for doping
the VBS state, and the analogous theory for the sF state appears in
Section~\ref{sf2}.

\section{Doping the Valence Bond Solid}
\label{qdn}

An explicit microscopic theory of the doping of the VBS state
described in Section~\ref{vbs1} has already appeared in
Ref.~\cite{psgdimers}. There we realized the VBS state by a quantum
dimer model, and then carried out a detailed duality transformation
on the doped quantum dimer model. Here, we will show how the same
results can be obtained by more abstract, but also more general,
symmetry arguments based upon the PSG. The present analysis makes it
clear that the results of Ref.~\cite{psgdimers} are applicable to a
far more general class of models, and also allows us to obtain new
results on the nature of superfluid-insulator transitions.

\subsection{Undoped insulator}
\label{qdn1}

The theory of the insulator was outlined in Section~\ref{vbs1}, and
has been discussed in much detail elsewhere. Here, we focus
exclusively on the crucial PSG properties. First, we complete the
mappings in Eq.~(\ref{ta1}) to the full set of symmetry operations
for the underlying antiferromagnet. We will consider the following
operations here, and in the remainder of the paper
\begin{eqnarray}
T_x&:&\mbox{Translation along the $x$ axis by one lattice
site.} \nonumber \\
T_y&:&\mbox{Translation along the $y$ axis by one lattice
site.} \nonumber \\
R_{\pi/2}^{\rm dual}&:&\mbox{Rotation by $90^\circ$ about a dual lattice site;} \nonumber \\
&~&\mbox{with the origin on a dual lattice site,} \nonumber \\
&~&\mbox{$x \rightarrow y$ and $y \rightarrow -x$.} \nonumber \\
I_{x}^{\rm dual}&:&\mbox{Reflection about the $y$ axis of the dual lattice;} \nonumber \\
&~&\mbox{with the origin on a dual lattice site,} \nonumber \\
&~&\mbox{$x \rightarrow -x$ and $y \rightarrow y$.} \nonumber \\
\mathcal{T}&:&\mbox{Time reversal.} \label{defpsg}
\end{eqnarray}
The mappings of $\mathcal{A}_\mu$ are easily determined from Eqs.
(\ref{nls1}) and (\ref{nls2}), and we obtain
\begin{eqnarray}
T_x&:& \mathcal{A}_\mu \rightarrow - \mathcal{A}_\mu \nonumber \\
T_y&:& \mathcal{A}_\mu \rightarrow - \mathcal{A}_\mu \nonumber \\
R_{\pi/2}^{\rm
dual}&:&\mathcal{A}_\tau\rightarrow-\mathcal{A}_\tau~;~\mathcal{A}_x\rightarrow-\mathcal{A}_y~;~
\mathcal{A}_y\rightarrow\mathcal{A}_x
\nonumber \\
I_{x}^{\rm
dual}&:&\mathcal{A}_\tau\rightarrow-\mathcal{A}_\tau~;~\mathcal{A}_x\rightarrow\mathcal{A}_x~;~
\mathcal{A}_y\rightarrow-\mathcal{A}_y
\nonumber \\
\mathcal{T}&:&\mathcal{A}_\tau \rightarrow
\mathcal{A}_\tau~;~\mathcal{A}_x \rightarrow
-\mathcal{A}_x~;~\mathcal{A}_y \rightarrow -\mathcal{A}_y.
\label{psgcalA}
\end{eqnarray}
We have not explicitly written out the transformations of the
spacetime co-ordinates of the fields above, because they are evident
in Eq.~(\ref{defpsg}).

The PSG of the monopoles in $\mathcal{A}_\mu$ requires careful
consideration of the Berry phases in Eq.~(\ref{berry1}). As shown by
Haldane, \cite{haldane} each monopole acquires a net phase factor,
which (for $S=1/2$) then leads to the following PSG transformations
\cite{senthil2} for the monopole annihilation operator $m$
\begin{eqnarray}
T_x&:& m \rightarrow i m^\dagger \nonumber \\
T_y&:& m \rightarrow -i m^\dagger \nonumber \\
R_{\pi/2}^{\rm dual}&:&m \rightarrow  m^\dagger
\nonumber \\
I_{x}^{\rm dual}&:& m \rightarrow m
\nonumber \\
\mathcal{T}&:&m \rightarrow m. \label{psgm}
\end{eqnarray}
As has been discussed elsewhere, \cite{senthil1,senthil2} $e^{i
\pi/4} m$ has the same transformation properties as the VBS order
parameter. The condensation of $m$ in the compact U(1) gauge theory
for the VBS phase then implies the appearance of VBS order in the
ground state.

\subsection{Doped antiferromagnet}
\label{sec:dope1}

We will consider the charged excitations of the doped
antiferromagnet in this subsection: the spin excitations will be
considered later in Section~\ref{sec:vbsferm}.

As noted in Section~\ref{dta}, we represent the charge carriers in
the doped antiferromagnet by $S=0$, charge $e$ holon degrees of
freedom $b_1 \sim e^{-i \theta_1}$ and $b_2 \sim e^{-i \theta_2}$.
The boson $b_1$ carries charge $+1$ under $\mathcal{A}_\mu$, while
the boson $b_2$ carries charge $-1$ under $\mathcal{A}_\mu$. The
total density of holes is $\delta$, and the density of each species
of boson is $\delta/2$.

We can now obtain an effective theory for these bosons, constrained
by their transformations under the PSG. Interestingly, there is
little arbitrariness in this PSG: it is almost entirely determined
by the PSG of $\mathcal{A}_\mu$ in Eq.~(\ref{psgcalA}), the
$\mathcal{A}_\mu$ charge assignments of the bosons above, and the
requirements of gauge invariance. We will actually not need the
explicit form of the boson action; just the PSG will suffice for our
subsequent duality mapping to the vortices. A simple analysis shows
that the PSG of the bosons is
\begin{eqnarray}
T_x&:& b_1 \rightarrow b_2~;~b_2 \rightarrow b_1 \nonumber \\
T_y&:& b_1 \rightarrow b_2~;~b_2 \rightarrow b_1  \nonumber \\
R_{\pi/2}^{\rm dual}&:&b_1 \rightarrow i b_2~;~b_2 \rightarrow i b_1
\nonumber \\
I_{x}^{\rm dual}&:& b_1 \rightarrow b_2~;~b_2 \rightarrow b_1
\nonumber \\
\mathcal{T}&:&b_1 \rightarrow b_1^\dagger~;~b_2 \rightarrow
b_2^\dagger . \label{psgb}
\end{eqnarray}
The factors of $i$ in the transformations under $R_{\pi/2}^{\rm
dual}$ are not determined by the requirements of gauge invariance.
Instead, as discussed in some detail in Ref.~\cite{psgdimers}, these
phase factors depend upon microscopic details, and the nature of the
short-range pairing in the superconducting ground states. The phase
factors displayed above are those appropriate to $d$-wave pairing.
In any case, such phase factors do not play any role in the duality
to the vortex degrees of freedom.

We proceed to apply the boson-vortex duality separately to both
species of bosons $b_{1,2}$. Then boson $b_1$ dualizes to a vortex
$\psi_1 \sim e^{-i \phi_1}$ and a non-compact U(1) gauge field
$A_{1\mu}$, while $b_2$ dualizes to a vortex $\psi_2 \sim e^{-i
\phi_2}$ and a non-compact U(1) gauge field $A_{2\mu}$. Again, it is
not necessary to write down the explicit form of the dual action as
the PSG will suffice in determining the low energy continuum limit
below. We can obtain the PSG of the dual fields by the requirements
of gauge invariance, the facts that the boson currents $\Delta_\mu
\theta_1 - \mathcal{A}_\mu$ and $\Delta_\mu \theta_2 +
\mathcal{A}_\mu$ must transform like the dual fluxes
$\epsilon_{\mu\nu\lambda} \Delta_\nu A_{1\lambda}$ and
$\epsilon_{\mu\nu\lambda} \Delta_\nu A_{2\lambda}$ respectively, and
conversely the dual vortex currents $\Delta_\mu \phi_1 - A_{1\mu}$
and $-\Delta_\mu \phi_2 + A_{2 \mu}$ must transform like the gauge
flux $\epsilon_{\mu\nu\lambda} \Delta_\nu \mathcal{A}_\lambda$. From
these requirements, it is not difficult to determine the PSG of the
dual vortex matter fields $\psi_{1,2}$
\begin{eqnarray}
T_x&:& \psi_1 \rightarrow \psi_2~;~\psi_2 \rightarrow \psi_1 \nonumber \\
T_y&:& \psi_1 \rightarrow \psi_2~;~\psi_2 \rightarrow \psi_1  \nonumber \\
R_{\pi/2}^{\rm dual}&:&\psi_1 \rightarrow \psi_2~;~\psi_2
\rightarrow \psi_1
\nonumber \\
I_{x}^{\rm dual}&:& \psi_1 \rightarrow \psi_2^\dagger~;~\psi_2
\rightarrow \psi_1^\dagger
\nonumber \\
\mathcal{T}&:&\psi_1 \rightarrow \psi_1~;~\psi_2 \rightarrow \psi_2,
\label{psgpsi}
\end{eqnarray}
and of the dual U(1) gauge fields $A_{1,2}$
\begin{eqnarray}
T_x&:& A_{1\mu} \rightarrow A_{2\mu}~;~A_{2\mu} \rightarrow A_{1\mu} \nonumber \\
T_x&:& A_{1\mu} \rightarrow A_{2\mu}~;~A_{2\mu} \rightarrow A_{1\mu} \nonumber \\
R_{\pi/2}^{\rm dual}&:& A_{1\tau} \rightarrow A_{2\tau}~;~A_{1x}
\rightarrow
A_{2y}~;~A_{1y} \rightarrow -A_{2x}~;~\nonumber \\
&~& A_{2\tau} \rightarrow A_{1\tau}~;~A_{2x} \rightarrow
A_{1y}~;~A_{2y} \rightarrow -A_{1x}
\nonumber \\
I_{x}^{\rm dual}&:& A_{1\tau} \rightarrow -A_{2\tau}~;~A_{1x}
\rightarrow
A_{2x}~;~A_{1y} \rightarrow -A_{2y}~;~\nonumber \\
&~& A_{2\tau} \rightarrow -A_{1\tau}~;~A_{2x} \rightarrow
A_{1x}~;~A_{2y} \rightarrow -A_{1y}
\nonumber \\
\mathcal{T}&:& A_{1\tau} \rightarrow -A_{1\tau}~;~A_{1x} \rightarrow
A_{1x}~;~A_{1y} \rightarrow A_{1y}~;~\nonumber \\
&~& A_{2\tau} \rightarrow -A_{2\tau}~;~A_{2x} \rightarrow
A_{2x}~;~A_{2y} \rightarrow A_{2y}. \label{psgA}
\end{eqnarray}

The PSG transformations in Eqs.~(\ref{psgm}), (\ref{psgpsi}), and
(\ref{psgA}) contain almost all the information needed to obtain the
vortex/monopole theory of the doped antiferromagnet. However, there
is an crucial ingredient that has been missing from our discussion
so far: the influence of the background density of $\delta/2$ of
each species of boson. In the dual vortex picture, this emerges as
an average background flux of both $A_{1,2}$ of $2 \pi (\delta/2)$
per plaquette. The consequences of such a background flux on the
vortex PSG were studied in some detail in Ref.~\cite{psgbosons}, and
we can simply transfer those results here separately to $\psi_1$ and
$\psi_2$. At a rational hole density in Eq.~(\ref{defpq}), the
$\psi_{1,2}$ vortices each generate $q$ degenerate low energy
fields. Following Ref.~\cite{psgbosons}, we denote these fields as
$\varphi_{1\ell}$ and $\varphi_{2\ell}$ where $\ell = 0,1 \ldots q$.
So at the moment there are a total of $2q$ vortex fields that
constitute the degrees of freedom of our low energy theory. The PSG
of the these fields follow from Eq.~(\ref{psgpsi}) and
Ref.~\cite{psgbosons}
\begin{eqnarray}
T_x&:& \varphi_{1\ell} \rightarrow \varphi_{2,\ell+1}~;~
\varphi_{2\ell} \rightarrow \varphi_{1,\ell+1} \nonumber \\
T_y&:& \varphi_{1\ell} \rightarrow \varphi_{2\ell} \omega^{-\ell}
~;~\varphi_{2\ell} \rightarrow \varphi_{1\ell} \omega^{-\ell}  \nonumber \\
R_{\pi/2}^{\rm dual}&:&\varphi_{1\ell} \rightarrow
\frac{1}{\sqrt{q}} \sum_{m=0}^{q-1} \varphi_{2m} \omega^{-\ell
m}~;~\nonumber \\
&~& \varphi_{2\ell} \rightarrow \frac{1}{\sqrt{q}} \sum_{m=0}^{q-1}
\varphi_{1m} \omega^{-\ell m}
\nonumber \\
I_{x}^{\rm dual}&:& \varphi_{1\ell} \rightarrow
\varphi_{2\ell}^\dagger~;~\varphi_{2\ell} \rightarrow
\varphi_{1\ell}^\dagger
\nonumber \\
\mathcal{T}&:&\varphi_{1\ell} \rightarrow
\varphi_{1\ell}~;~\varphi_{2\ell} \rightarrow \varphi_{2\ell},
\label{psgvarphi}
\end{eqnarray}
where the $\ell,m$ indices are all implicitly determined modulo $q$,
and
\begin{equation}
\omega \equiv e^{2 \pi i p/q}.
\end{equation}

In the gauge field sector, the low energy theory is most easily
expressed in terms of the uniform and staggered dual gauge fields
defined by
\begin{eqnarray}
A_{1\mu} &=& B_{s\mu} + B_{a \mu} \nonumber \\
A_{2\mu} &=& B_{s\mu} - B_{a \mu}. \label{defasa}
\end{eqnarray}

We are now ready to present the promised unification of the vortex
and monopole PSGs. Indeed, all results follow from the PSGs in
Eqs.~(\ref{psgm}) and (\ref{psgvarphi}), coupled with the
requirements of gauge invariance. Before proceeding, let us briefly
discuss the latter requirements. The $\varphi_{1\ell}$ fields have
charges of $+1$ and $+1$ under $B_{s \mu}$ and $B_{a \mu}$
respectively, while the $\varphi_{2\ell}$ fields have charges of
$+1$ and $-1$. The monopole changes the $\mathcal{A}_\mu$ gauge flux
by $2 \pi$, and we argued earlier in Section~\ref{dta} this implied
that a monopole transformed like $\psi_2^\dagger \psi_1$; in other
words, the monopole operator $m^\dagger$ has charges of $0$ and $+2$
under $B_{s \mu}$ and $B_{a \mu}$ respectively. These gauge charges
are summarized in Table~\ref{table1}.
\begin{table}
\begin{tabular}{||c||c|c|c|c|c|c||} \hline\hline
 & Vortex & Vortex & Monopole & Holon & Holon & Spinon \\
  & $\varphi_1$ & $\varphi_2$ & $m$ & $b_1$ & $b_2$ & $f_s$ \\
\hline\hline Dual & & & & & & \\
Gauge & 1 & 1 & 0 & --- & --- & --- \\
$B_{s\mu}$ &  &  &  & & & \\
\hline Dual & & & & & & \\
Gauge & 1 & -1 & 2 & --- & --- & --- \\
$B_{a\mu}$ & & & & & & \\
\hline
Direct & & & & & & $\eta_j$ (VBS) \\
Gauge & --- & --- & --- & 1 & -1 & and \\
$\mathcal{A}_\mu$ &  &  & & & & 1 (sF) \\
\hline\hline
\end{tabular}
\caption{Assignments of charges under the various gauge field. The
direct gauge field $\mathcal{A}_\mu$ is related to spin singlet
fluctuations in the insulating RVB state, as discussed in
Section~\ref{sec:intro}. The flux of the dual gauge field $B_{s\mu}$
measures the electrical supercurrent. The dual gauge field $B_{a
\mu}$ is the Chern-Simons dual of $\mathcal{A}_\mu$ {\em i.e.\/} the
flux of $B_{a \mu}$ measures the current associated with
$\mathcal{A}_\mu$, and vice versa.} \label{table1}
\end{table}

\subsection{Low energy theory}
\label{sec:lw1}

We are now ready to begin presentation of the effective action. This
has the structure
\begin{equation}
\mathcal{S} = \int d^2 x d \tau \left[\mathcal{L}_f
[\mathcal{A}_\mu] + \mathcal{L}_{\rm U(1)} + \mathcal{L}_{2 \varphi}
+ \mathcal{L}_{4 \varphi} + \mathcal{L}_m \right] \label{vbsaction}
\end{equation}

The first term, $\mathcal{L}_f [\mathcal{A}_\mu]$, is the action for
the $S=1/2$ fermionic spinons; we defer consideration of this term
to Section~\ref{sec:vbsferm}.

The second term, $\mathcal{L}_{\rm U(1)}$ is the action for the U(1)
gauge fields, $\mathcal{A}_\mu$, $B_{s\mu}$, and $B_{a\mu}$. Apart
from the conventional Maxwell terms, there is also a Chern-Simons
term which ensures that the $\mathcal{A}_\mu$ flux is equal to the
vortex current associated with $A_{1\mu} - A_{2 \mu}$: this is
demanded by the discussion in Section~\ref{dta} which showed that
there is $\pm \pi$ $\mathcal{A}_\mu$ flux associated with the
$\psi_{1,2}$ vortices.
\begin{eqnarray}
\mathcal{L}_{\rm U(1)} &=& \frac{K}{2} \left(
\epsilon_{\mu\nu\lambda}
\partial_\nu \mathcal{A}_\lambda \right)^2 + \frac{K_s}{2} \left( \epsilon_{\mu\nu\lambda}
\partial_\nu B_{s \lambda} \right)^2 \nonumber \\ &~&~~~~~~~~~~~~~~+ \frac{K_a}{2} \left( \epsilon_{\mu\nu\lambda}
\partial_\nu B_{a \lambda} \right)^2 + \frac{i}{\pi}
\epsilon_{\mu\nu\lambda} \partial_\mu B_{a \nu}
\mathcal{A}_{\lambda} .\label{la}
\end{eqnarray}
Further $\mathcal{A}_\mu$ dependence in the action appears only in
the fermionic term $\mathcal{L}_f [\mathcal{A}_\mu]$. If the
fermions are gapped, they can be safely integrated out, at the cost
of a renormalization of the coupling $K$. In this case, we can also
integrate out $\mathcal{A}_\mu$ from Eq.~(\ref{la}), and conclude
that the $B_{a \mu}$ gauge field is gapped and can be dropped from
further considerations. The situation with gapless fermions is far
more complicated, and requires a full analysis of the coupled
$\mathcal{A}_\mu$, $B_{a \mu}$, and the fermionic excitations.

The action $\mathcal{L}_{2 \varphi}$ is the quadratic term for the
vortex fields $\varphi_{1,2}$ consistent with the gauge charge
assignments and the PSG. After appropriate rescalings of the
$\varphi_{1,2}$ fields, and rescaling of time and space
co-ordinates, this has the form
\begin{eqnarray}
\mathcal{L}_{2 \varphi} &=& \sum_{\ell=0}^{q-1} \Bigl\{
\left|\left(\partial_\mu - i B_{s \mu} - i B_{a \mu}
\right)\varphi_{1\ell}\right|^2 + \left|\left(\partial_\mu - i B_{s
\mu} + i B_{a \mu} \right)\varphi_{2\ell}\right|^2 \nonumber
\\ &~&~~~~~~+ s \left( |\varphi_{1\ell}|^2 + |\varphi_{2\ell}|^2 \right)
\Bigr\} + \widetilde{\mathcal{L}}_{2 \varphi}. \label{l2a}
\end{eqnarray}
Here we displayed the standard kinetic terms which identifies the
$2q$ $\varphi_{1,2\ell}$ fields as relativistic complex scalars. The
``mass'' $s$ is the tuning parameter which accesses various phases
of the doped antiferromagnet. For large positive $s$, the vortices
are gapped, and the ground state is superconducting. Condensates of
vortices or vortex-anti-vortex pairs can form a lower doping,
leading to insulating or supersolid phases.

There is an additional quadratic term, denoted
$\widetilde{\mathcal{L}}_{2 \varphi}$ above, invariant under all PSG
transformations in Eq.~(\ref{psgvarphi}, which is allowed in some
cases. For the special values $q=4n+2$ ($n$ integer), it is easy to
check that
\begin{equation}
\widetilde{\mathcal{L}}_{2 \varphi} = i \Xi \sum_{\ell=0}^{q-1}
(-1)^\ell \left( \varphi_{1\ell}^\ast \varphi_{1,\ell+q/2} -
\varphi_{2\ell}^\ast \varphi_{2,\ell+q/2} \right) \label{l2p}
\end{equation}
is allowed, where $\Xi$ is a real coupling constant. More
physically, this term has its origin on a staggered potential that
can act on the $b_{1,2}$ holons. As derived explicitly in
Ref.~\cite{psgdimers}, or by the requirements of the PSG in
Eq.~(\ref{psgb}), the boson Hamiltonian can contain the term
\begin{displaymath}
\sum_j \eta_j \left( b^{\dagger}_{1j} b_{1j} - b^{\dagger}_{2j}
b_{2j} \right).
\end{displaymath}
Upon dualizing to the vortices, this term appears as a staggered
flux acting on the vortices, which in the continuum limit generates
Eq.~(\ref{l2p}).

The term $\mathcal{L}_{4 \varphi}$ is a quartic polynomial in the
vortex fields $\varphi_{1,2}$. This polynomial is constrained by the
PSG transformations in Eq.~(\ref{psgvarphi}), in a manner that has
been discussed at some length in Ref.~\cite{psgbosons}. These
quartic terms control the precise structure of the density-wave
order in the supersolid and insulating phases.

Finally, we turn to the main term being introduced in this paper:
the $\mathcal{L}_{m}$ which couples the vortices and the monopoles.
In obtaining this term, we simply have to search for invariants
under the PSGs in Eq.~(\ref{psgm}) and (\ref{psgvarphi}). A
straightforward analysis shows that the lowest order invariant,
present only for $q=2n$ is
\begin{eqnarray}
\mathcal{L}_m  = && \lambda m^\dagger \sum_{\ell=0}^{q-1} \Biggl[
e^{- i \pi/4} (-1)^\ell \varphi_{1 \ell}^\ast \varphi_{2 \ell} +
e^{i \pi /4} \varphi_{1 \ell}^{\ast} \varphi_{2,\ell + q/2} \Biggr]
\nonumber
\\ && + \mbox{c.c.}, \label{lm}
\end{eqnarray}
where $\lambda$ is a complex coupling constant. Note that this is
the only term which couples together the $\varphi_1$ and $\varphi_2$
vortices, and so places a crucial role in determining the vortex
spectrum.

\subsection{Vortex spectrum in the superconductor}
\label{vortexsup}

The vortices are gapped in the superconductor, and here we are
interested in the nature of the single vortex excited state.
(Strictly speaking, a single vortex has a logarithmically divergent
energy, and so the excitations are vortex-anti-vortex pairs---we imagine
there is an anti-vortex far away, and examine the motion of the
vortex). Once a $\varphi_1$ vortex has been created (say), it can
transmute into a $\varphi_2$ vortex by the monopole tunnelling
process in Eq.~(\ref{lm}). In the superconductor, we expect that the
flux associated with this monopole will be largely confined to the
vortex cores, and so the action for the monopole will be finite.
Under these circumstances, it is legitimate to simply treat the
field $m$ as a constant \cite{senthil2} (in the absence of gapless
fermions, it is also valid to replace $m$ by a constant across the
transition to the insulator).

The spectrum of a single vortex is then simply determined by
diagaonalizing the quadratic action $\mathcal{L}_{2 \varphi} +
\mathcal{L}_m$ and ignoring the gauge fields. We describe the
results of this procedure below for different values of $q$ in the
following subsections.

\subsubsection{$q =1,3~\mbox{(mod 4)}$}

For $q$ odd, $\mathcal{L}_m=0$ and
$\widetilde{\mathcal{L}}_{2\varphi}=0$, and so the $\varphi_{1\ell}$
and $\varphi_{2\ell}$ remain vortex eigenstates. The degeneracy of
vortex states for this case is therefore $q' = 2 q$.

\subsubsection{$q =0~\mbox{(mod 4)}$}

Next, we consider the case where $q$ is a multiple of 4. Now
$\mathcal{L}_m$ in Eq.~(\ref{lm}) is non-zero, and it splits the
$2q$ dimensional vortex space into $4 \times 4$ blocks in which
$\varphi_{1\ell}$, $\varphi_{1,\ell+q/2}$, $\varphi_{2\ell}$, and
$\varphi_{2,\ell+q/2}$ are coupled to each other. In each block, the
action has the structure
\begin{equation}
\Omega^2 + k^2 +
s + \left( \begin{array}{cccc} 0 & 0 & \alpha & \beta \\
0 & 0 & \beta & \alpha  \\
\alpha^\ast & \beta^\ast & 0 & 0 \\
 \beta^\ast & \alpha^\ast & 0 & 0
\end{array} \right)
\label{q4n}
\end{equation}
where $\alpha \equiv \lambda (-1)^\ell m^\dagger e^{-i \pi/4}$,
$\beta \equiv \lambda m^{\dagger} e^{i \pi/4}$, $\Omega$ is the
frequency, and $k$ is the wavevector. The eigenvalues of this action
are $ \pm \left( |\alpha|^2 + |\beta|^2 \pm (\alpha^\ast \beta +
\beta^\ast \alpha) \right)^{1/2} + \Omega^2 + k^2 + s$. Now note
that $\alpha^\ast \beta + \beta^\ast \alpha = 0$, and hence there
are 2 pairs of doubly degenerate eigenvalues. Consequently, the $2q$
vortex states have split into 2 blocks of $q$ degenerate states. The
lowest energy vortex states therefore have the degeneracy $q' = q$.

\subsubsection{$q = 2~\mbox{(mod 4)}$}

Finally, consider the remaining case $q=4n+2$, where $n$ is a
positive integer. Now $\widetilde{\mathcal{L}}_{2 \varphi}$ in Eq.
(\ref{l2p}) is also non-zero, but the action still has a $4 \times
4$ block structure. The matrix in Eq. (\ref{q4n}) is replaced by
\begin{equation}
\Omega^2 + k^2 +
s + \left( \begin{array}{cccc} 0 & i \gamma & \alpha & \beta \\
-i \gamma & 0 & \beta & -\alpha  \\
\alpha^\ast & \beta^\ast & 0 & -i \gamma \\
 \beta^\ast & -\alpha^\ast & i \gamma & 0
\end{array} \right)
\label{q4n2}
\end{equation}
where $\gamma \equiv (-1)^{\ell} \Xi$. Now the 4 eigenvalues are
$\rho_1 \left( |\alpha|^2 + |\beta|^2 - i \rho_2 (\alpha^\ast \beta
- \beta^\ast \alpha) \right)^{1/2} + \Omega^2 + k^2 + s + \rho_2
\gamma$, where $\rho_{1,2} = \pm 1$. All 4 eigenvalues are distinct,
and so the $2q$ vortex eigenstates have now split into blocks of
$q/2$ degenerate states. The lowest energy vortex states therefore
have the degeneracy $q' = q/2$.

\subsubsection{General discussion}

We summarize the above results on the degeneracy of the vortex
states in Table~\ref{table2}.
\begin{table}
\begin{tabular}{||c|c||} \hline\hline
$q$ & $q'$ \\
\hline\hline ~$2n+1$~ & $2q$ \\
\hline
$4n$ & $q$ \\
\hline
$4n+2$ & ~$q/2$~ \\
\hline \hline
\end{tabular}
\caption{Degeneracy of the vortex states, $q'$, at hole density
$\delta$. Here $\delta$ is related to $q$ as in Eq.~(\ref{defpq}),
and $n$ is a positive integer.} \label{table2}
\end{table}
The numerology of the vortex degeneracy seems rather mysterious, but
actually has a simple physical interpretation. The results in
Table~\ref{table2} can be reproduced by the simple formula
\begin{equation}
\frac{p'}{q'} = \frac{1}{2} - \frac{p}{q}, \label{defqprime}
\end{equation}
where $p'$ and $q'$ are relatively prime integers (as are $p$ and
$q$). Recalling that $p/q$ is half the density of holes
(Eq.~(\ref{defpq})) away from the half-filled Mott insulator, we now
see that $p'/q'$ is half the density of electrons {\em i.e.\/}
$p'/q'$ is the total density of Cooper pairs per unit cell. So the
degeneracy of the vortex spectrum is always equal to that of a model
of elementary bosons with the boson density equal to the density of
Cooper pairs. This is one of the main results of this paper.

The above result can also be understood directly from the PSG.
Consider the action of the PSG on field combinations which are
neutral under $B_{a\mu}$ (only such combinations appear in the
action above). For this, we define
\begin{equation}
\overline{\varphi}_{1\ell} = \varphi_{1\ell} \, m^{1/2} ~~;~~
\overline{\varphi}_{2\ell} = \varphi_{2\ell} \left( m^{\dagger}
\right)^{1/2} \label{defphib}
\end{equation}
The action of the PSG on these fields is
\begin{eqnarray}
T_x&:& \overline{\varphi}_{1\ell} \rightarrow e^{i \pi/4}
\overline{\varphi}_{2,\ell+1}~;~\overline{\varphi}_{2\ell}
\rightarrow e^{-i \pi/4} \overline{\varphi}_{1,\ell+1} \nonumber \\
T_y&:& \overline{\varphi}_{1\ell} \rightarrow e^{-i \pi/4}
\overline{\varphi}_{2\ell}
\omega^{-\ell}~;~\overline{\varphi}_{2\ell} \rightarrow
e^{i \pi/4} \overline{\varphi}_{1\ell}\omega^{-\ell} \nonumber \\
R_{\pi/2}^{\rm dual}&:& \overline{\varphi}_{1\ell} \rightarrow
\frac{1}{\sqrt{q}} \sum_{m=0}^{q-1} \overline{\varphi}_{2m}
\omega^{-\ell m}~;~\nonumber \\
&~& \overline{\varphi}_{2\ell} \rightarrow \frac{1}{\sqrt{q}}
\sum_{m=0}^{q-1} \overline{\varphi}_{1m} \omega^{-\ell m} \label{e7}
\end{eqnarray}
Note that the transformations in Eq.~(\ref{e7}) obey
\begin{equation}
T_x T_y = -\omega T_y T_x = \exp \left( -\frac{2 \pi i p'}{q'}
\right) T_y T_x,
\end{equation}
thus explaining the $q'$-fold degeneracy of vortex states.

\subsection{Fermionic excitations}
\label{sec:vbsferm}

The analysis of Ref.~\cite{psgdimers} of spin $S=1/2$, neutral
excitations of the quantum dimer and related models applies directly
to our theory of the doped VBS phase. These excitations are
represented by Fermi operators $f_{j\sigma}$ on the sites, $j$, of
the square lattice, with  $\sigma = \uparrow, \downarrow$ a spin
index. This fermion carries $\mathcal{A}_\mu$ gauge charge $\eta_j$,
as indicated in Table~\ref{table1}. Also crucial are its
transformation properties under the PSG
\begin{eqnarray}
T_x&:& f_{\sigma} \rightarrow f_{\sigma}\nonumber \\
T_y&:& f_{\sigma} \rightarrow f_{\sigma}\nonumber \\
R_{\pi/2}^{\rm dual} &:& f_{\sigma} \rightarrow if_{\sigma}\nonumber \\
I_x^{\rm dual} &:& f_{\sigma} \rightarrow f_{\sigma}\nonumber \\
\mathcal{T}&:& f_\sigma \rightarrow \epsilon_{\sigma\sigma'}
f^{\dagger}_{\sigma'}~~;~~f_\sigma^\dagger \rightarrow
-\epsilon_{\sigma\sigma'} f_{\sigma'}, \label{psgf}
\end{eqnarray}
where $\epsilon_{\sigma \sigma'}$ is the antisymmetric tensor. The
transformations of $f^\dagger_{\sigma}$ are the Hermitean conjugates
of those of $f_\sigma$, except for the case of time-reversal.
Time-reversal does not correspond to a canonical unitary
transformation, and we have chosen to describe time-reversal as a
symmetry of the Grassman coherent-state path integral. The minus
sign in the time-reversal transformation of $f^\dagger$ above is
allowed since the $f^\dagger$ and $f$ fields are {\sl not} complex
conjugates but actually independent in the Grassman integral.  The
factor $i$ in the transformation under $R_{\pi/2}^{\rm dual}$ is
related to the corresponding factor in the PSG for the holons in
Eq.~(\ref{psgb}). It corresponds to an assumption of $d$-wave
pairing.

With the gauge charges and the PSG at hand, we can write down
allowed terms in the fermionic contribution, $\mathcal{L}_f
[\mathcal{A}_\mu]$, to the action:
\begin{eqnarray}
\mathcal{L}_f [\mathcal{A}_\mu] &=& \sum_j f_{j\sigma}^{\dagger}
\left(\frac{\partial }{\partial \tau} - i \eta_j \mathcal{A}_{j
\tau} \right) f_{j \sigma} + v\sum_{j} f_{j\sigma}^\dagger
  f_{j\sigma}^{\vphantom\dagger} \nonumber  \\ &+& \sum_{j\alpha} \Delta_\alpha
  e^{-i\eta_j \mathcal{
      A}_{j\alpha}} f_{j\sigma}^\dagger \epsilon_{\sigma \sigma'}
  f_{j+\alpha,\sigma'}^\dagger + {\rm H.c.} \nonumber \\ &-& t \sum_{j,\alpha=x,y} b_{j 1} b_{j+\alpha,2}^\dagger
f_{j\sigma}^\dagger f_{j+\alpha,\sigma}  + {\rm H.c.} \label{eq:clf}
\end{eqnarray}
Clearly, $\Delta_\alpha$ is a pairing amplitude, and we have
$\Delta_x=-\Delta_y$.

In principle, it is now possible to take the continuum limit of
$\mathcal{L}_f$ and then examine the properties of
Eq.~(\ref{vbsaction}). However, with the expected proliferation of
monopoles in the insulator, we expect that $\mathcal{A}_\mu$
fluctuations are very strong, and it may be more appropriate to use
fields that are neutral under $\mathcal{A}_\mu$. The quadratic terms
in the fermion dispersion in Eq.~(\ref{eq:clf}) have a gapped
fermion spectrum, and so no low energy fermions are available to
suppress monopole events.

Rather, it seems more reasonable to assume that strong
$\mathcal{A}_\mu$ fluctuations bind the $b_{1,2}$ holons to the
$f_\sigma$ fermions. We can then express the fermionic Hamiltonian
in terms of charge $e$ physical electron operators $c_{j\sigma}$:
\begin{eqnarray}
c_{j\sigma} &=& e^{i \theta_{j1} } f_{j\sigma}~~\mbox{for $\eta_j =
1$} \nonumber \\
c_{j\sigma} &=& e^{i \theta_{j2} } f_{j\sigma}~~\mbox{for $\eta_j =
-1$}
\end{eqnarray}
At the same time, we can work with the $\overline{\varphi}_{1,2}$
fields defined in Eq.~(\ref{defphib}), and then all degrees of
freedom are neutral under the $\mathcal{A}_\mu$ and $B_{a\mu}$ gauge
fields: the action is expressed entirely in terms of the
$c_{j\sigma}$, $\overline{\varphi}_{1,2}$, and $B_{s\mu}$ fields.

The Hamiltonian for the $c_{j \sigma}$ fermions is essentially
identical to the BCS Hamiltonian
\begin{eqnarray}
\mathcal{L}_c &=& \sum_j c_{j \sigma}^\dagger \frac{\partial c_{j
\sigma}}{\partial \tau} +  v\sum_{j} c_{j\sigma}^\dagger
  c_{j\sigma}^{\vphantom\dagger} \label{BCS} \\ &+& \sum_{j\alpha} \Delta_\alpha
  e^{-i(\theta_{j1} + \theta_{j+\alpha,2} +  \mathcal{
      A}_{j\alpha})} c_{j\sigma}^\dagger \epsilon_{\sigma \sigma'}
  c_{j+\alpha,\sigma'}^\dagger + {\rm H.c.}
  -t \sum_{j,\alpha=x,y} c_{j\sigma}^\dagger c_{j+\alpha,\sigma} +
{\rm H.c.}\nonumber
\end{eqnarray}
Notice that the exponential in the second line containing a
gauge-invariant combination of fields which carries physical
electrical charge 2. So the $c_{\sigma}$ fermions couple only to the
physical BCS order parameter. Further, just as in the BCS theory,
the $c_\sigma$ fermions display the usual gapless nodal fermion
excitations of a $d$-wave superconductor.

At this point, the coupling of the $c_{\sigma}$ fermions to the
vortices can be described following the methods discussed in a
variety of papers in the literature
\cite{ft,ashwin,sf,bf,grsf,predrag}: it yields a theory for neutral
fermions valid in a regime where the fermionic excitations can be
gapless. After a singular gauge transformation which eliminates the
phase of the pairing amplitude \cite{ft,ashwin}, the vortices have
two important effects of the quasiparticle motion: the fermions
acquire a `statistical' phase of $\pi$ upon encircling a
$\overline{\varphi}_{1,2}$, and also acquire a ``Doppler shift''
proportional to the local superflow velocity. The statistical phase
is implemented by coupling both the vortices and the fermions to
U(1) gauge fields, $\alpha_\mu$, and $a_\mu$ respectively, along
with a mutual Chern-Simons term
\begin{equation}
\mathcal{L}_{cs} [ \alpha_\mu, a_\mu] = \frac{i}{\pi}
\epsilon_{\mu\nu\lambda} a_\mu \partial_\nu \alpha_\lambda.
\label{lcs}
\end{equation}
These gauge fields have to couple to conserved currents, and for the
fermions a convenient choice\cite{ft,ashwin,bfn} is the $z$
component of the spin. We denote the resulting fermionic Lagrangian
by $\mathcal{L}_c [a_\mu]$, and refer the reader to these earlier
works for the explicit form. For our purposes, we need the
transformations of the $a_\mu$ gauge field under the square lattice
symmetry operations. Using the usual transformations of the electron
operator $c_{\sigma}$ under the square lattice symmetry operations,
and with requirements of gauge invariance, we can easily deduce the
following PSG for $a_\mu$:
\begin{eqnarray}
T_x&:& a_\mu \rightarrow  a_\mu \nonumber \\
T_y&:& a_\mu \rightarrow  a_\mu \nonumber \\
R_{\pi/2}^{\rm dual}&:&a_\tau\rightarrow a_\tau~;~a_x\rightarrow
a_y~;~ a_y\rightarrow -a_x
\nonumber \\
I_{x}^{\rm dual}&:&a_\tau\rightarrow a_\tau~;~a_x\rightarrow -a_x~;~
a_y\rightarrow a_y
\nonumber \\
\mathcal{T}&:&a_\tau \rightarrow -a_\tau~;~a_x \rightarrow a_x~;~a_y
\rightarrow  a_y. \label{psga}
\end{eqnarray}
For the vortex sector, we couple $\alpha_\mu$ to the vortex current.
We assume that a diagonalization of the vortex spectrum has been
carried out as discussed in Section~\ref{vortexsup}, and focus
exclusively on the $q'$ degenerate low energy vortex modes. We
choose these modes as eigenvectors of the $T_y$ operator, and denote
these modes simply as $\overline{\varphi}_m$, with the index
$m=0,1,\dots (q'-1)$. Collecting these terms, our theory for the
vicinity of the superconductor-insulator transition is then
\begin{eqnarray}
&~& \mathcal{L}_{{\rm dSC-VBS},\overline{\varphi}} = \frac{K_s}{2}
\left( \epsilon_{\mu\nu\lambda} \partial_\nu B_{s \lambda} \right)^2
+ \mathcal{L}_{cs}[\alpha_\mu, a_\mu] \nonumber
\\ &+& \sum_{m=0}^{q'-1} \Bigl\{ \left|\left(\partial_\mu - i B_{s
\mu} - i \zeta_m \alpha_{\mu} \right)\overline{\varphi}_{m}\right|^2
+  s |\overline{\varphi}_{m}|^2 \Bigr\}+ \mathcal{L}_{4
\overline{\varphi}} + \mathcal{L}_c [a_\mu]. ~~\label{dSCVBS}
\end{eqnarray}
The action for the $B_{s\mu}$ field has been written in a schematic
relativistic form, which is appropriate for short-range interactions
between the bosons---the Coulomb interactions lead to modifications
presented in Ref.~\cite{predrag}. Note that the $\mathcal{A}_\mu$,
$B_{a\mu}$, and $m$ fields have dropped out. The term
$\mathcal{L}_{4\overline{\phi}}$ now includes quartic invariants in
the $\overline{\varphi}_m$ which are invariant under Eq.~(\ref{e7}).

The $\overline{\varphi}_m$ fields have charges $\zeta_m$ under the
gauge field $\alpha_\mu$. If $\alpha_\mu$ was coupling to the total
vortex current, then we should choose all $\zeta_m = 1$. However,
this gauge field implements only a statistical phase factor of -1,
and this can be obtained by choosing arbitrary odd integer
$\zeta_m$.

An action closely related to $\mathcal{L}_{{\rm
dSC-VBS},\overline{\varphi}}$ has been examined previously by
Lannert {\em et al.\/} \cite{lfs} for the case of Dirac spectrum in
$\mathcal{L}_c [a_\mu]$. They considered a simplified model at
half-filling with $q'=2$. As discussed by Lannert {\em et al.\/},
the Chern-Simons term in $\mathcal{L}_{{\rm
dSC-VBS},\overline{\varphi}}$ drives confinement in the insulating
phase with $\langle \overline{\varphi} \rangle \neq 0$: each $\Psi$
fermion has a $\alpha_\mu$ flux tube attached to it, and this
acquires an additional $B_{s\mu}$ flux tube when the total flux
coupled to the $\overline{\varphi}$ is expelled; the latter implies
attachment of charge $e$ to the spinons, and a likely gapping of the
fermion spectrum in the insulator. A notable feature of this theory
is that the nodal fermions remain gapless all the way up to the
critical point. A full PSG analysis on the direct lattice seems
necessary to verify such a scenario (as in Section III.B of
Ref.~\cite{psgbosons}), but below we present some plausible
constraints under which such a critical point may obtain.

We close this subsection by discussing some issues related to the
PSG properties of $\mathcal{L}_{{\rm dSC-VBS},\overline{\varphi}}$.
While the its properties should be invariant under arbitrary choices
for $\zeta_m$, our approximate analysis in the following subsection
will find dependence on the values of $\zeta_m$. It is therefore
useful to find a choice of $\zeta_m$ in which the invariance under
the PSG is as explicit as possible, without using non-perturbative
properties of the Chern-Simons term. We will see in the following
subsection that it is useful to satisfy the constraint $\sum_m
\zeta_m = 0$. This is clearly not possible for $q'$ odd (which
corresponds to $q=4n+2$), and so we will not consider this case
further. For other values of $q$, we divide the vortex fields into 2
sets, one with $\zeta_m=1$ and the other with $\zeta_m=-1$.
Invariance under the PSG is then possible provided the set of fields
with $\zeta_m=1$ either transform only among themselves, or all
transform into fields with $\zeta_m=-1$ (and conversely for the
fields with $\zeta_m=-1$). For $q$ odd (with $q'=2q$), there is no
monopole-induced mixing between the two sets of
$\overline{\varphi}_{1\ell}$ and the $\overline{\varphi}_{2\ell}$
fields, and we can simply assign $\zeta_m=1$ for the first set, and
$\zeta_m=-1$ for the second set. For $q$ a multiple of 4 (with
$q'=q$), we consider explicitly the {\em permutative\/} PSG for
$q=4$. This is specified in Eq.~(C5) of Ref.~\cite{psgbosons}, and
we see that, in this permutative vortex basis, the needed conditions
are satisfied for $\zeta_0=\zeta_2=1$ and $\zeta_1=\zeta_3=-1$. We
suspect a similar choice is possible for other $q$ multiples of 4.
In all the cases for which suitable $\zeta_m$ are possible, the PSG
of the $\overline{\varphi}_m$ implies a corresponding PSG for the
$\alpha_\mu$, which we will not write explicitly. Combining this
with the PSG for the $a_\mu$ in Eq.~(\ref{psga}), we finally have to
test the invariance of the Chern-Simons term $\mathcal{L}_{cs}
[\alpha_\mu, a_\mu]$ in Eq.~(\ref{lcs}): we find that
$\mathcal{L}_{cs} [\alpha_\mu, a_\mu]$ changes sign under most PSG
transformations. However, this sign is clearly not physically
significant because statistical phases of $\pm \pi$ are equivalent.

\subsection{`Undualizing' to fractionally charged bosons}
\label{sec:undual}

We now examine the critical theory in Eq.~(\ref{dSCVBS}), associated
with the condensation of the $\overline{\varphi}$ vortices, by
undoing the duality into direct lattice degrees of freedom. Here it
is possible to apply recent ideas \cite{senthil1,senthil2} on
`deconfined criticality' to `undualize' $\mathcal{L}_{{\rm
dSC-VBS},\overline{\varphi}}$ and obtain a theory expressed in terms
of fractionalized direct lattice degrees of freedom. The procedure
for doing this was discussed at length in Ref.~\cite{psgbosons} for
boson models at arbitrary rational filling on the square lattice.
Here, we can apply exactly the same procedure to the $q'$ vortices
$\overline{\varphi}_m$. As discussed in Section III.A of
Ref.~\cite{psgbosons}, such a transformation to direct lattice
bosons is possible only if it is possible to find a `permutative
representation' of the PSG. In the following, we assume that it is
possible to transform the relevant terms in $\mathcal{L}_{4
\overline{\varphi}}$ so that such a permutative representation
exists. Then, as shown in Section III.A of Ref.~\cite{psgbosons},
the $q'$ vortex fields $\overline{\varphi}_m$ and the U(1) gauge
field $B_{s\mu}$ can be `undualized' into $q'$ boson fields $\xi_m$
and $q'$ non-compact U(1) gauge fields
$\widetilde{\mathcal{A}}^m_\mu$. The $\xi_m$ boson fields each carry
physical electromagnetic charge $2e/q'$, and so each Cooper pair has
fractionalized into $q'$ elementary bosons. One contribution to the
direct formulation of Eq.~(\ref{dSCVBS}) is the theory of these
fields presented in Eq. (3.5) of Ref.~\cite{psgbosons}. Combining
these terms with the contribution of the Dirac fermions, we can
tentatively propose the following `undualized' formulation of
$\mathcal{L}_{{\rm dSC-VBS},\overline{\varphi}}$:
\begin{eqnarray}
&& \mathcal{L}_{{\rm deconfined}} = \sum_{m=0}^{q'-1} \left[\left|
\left(
\partial_{\mu} - i \widetilde{\mathcal{A}}^{m}_{\mu}  \right)
\xi_{m} \right|^2 + \widetilde{s} |\xi_m |^2 \right]  \nonumber \\
&+& \sum_{m,n} K_{mn} \left( \epsilon_{\mu\nu\lambda}
\partial_{\nu} \widetilde{\mathcal{A}}^{m}_{\lambda} \right)  \left( \epsilon_{\mu\rho\sigma}
\partial_{\rho} \widetilde{\mathcal{A}}^{n}_{\sigma} \right)
+ \sum_{m,n} \widetilde{v}_{mn} |\xi_m|^2 |\xi_n|^2 +\mathcal{L}_c
[a_\mu], \label{moose2}
\end{eqnarray}
where the couplings $K_{mn}$ and $v_{mn}$ can have the most general
form consistent with the permutative PSG. Now $\widetilde{s}$ is the
tuning parameter, which yields superducting ground states for
negative values of $\widetilde{s}$ where the $\xi_m$ bosons are
condensed, and insulating states for positive $\widetilde{s}$ where
the bosons are gapped. A specific duality transformation applied to
$\mathcal{L}_{{\rm dSC-VBS},\overline{\varphi}}$, along the lines of
Ref.~\cite{psgbosons}, does indeed yield precisely the action in
Eq.~(\ref{moose2}), with no residual Chern-Simons term---this
absence is one of the principal advantages of the direct
formulation. This duality transformation also shows that the $q'$
gauge fields $\widetilde{\mathcal{A}}^m_\mu$ and the $a_\mu$ gauge
field are not all independent, but certain linear combinations are
`Higgsed' out and so become gapped. We assume that we are working at
energies much lower than this energy gap, and so impose constraints
projecting out these linear combinations. In this manner we find the
constraints
\begin{eqnarray}
\sum_{m=0}^{q'-1} \widetilde{\mathcal{A}}^m_\mu &=& 0 \label{gaugeconstraint1} \\
a_\mu &=& \frac{1}{2} \sum_{m=0}^{q'-1} \zeta_m
\widetilde{\mathcal{A}}^m_\mu . \label{gaugeconstraint2}
\end{eqnarray}
The theory in Eq.~(\ref{moose2}) of $q'$, charge $2e/q'$,
relativistic complex scalars $\xi_m$ and the Dirac fermions $\Psi$,
coupled to the U(1) gauge field $a_\mu$ and the $q'$ U(1) gauge
fields $\widetilde{\mathcal{A}}^m_\mu$ (subject to the constraints
in Eqs.~(\ref{gaugeconstraint1}) and (\ref{gaugeconstraint2})), is
then our final theory for the deconfined criticality between a
$d$-wave superconductor and VBS insulators at Cooper pair density
$p'/q'$.

An important ingredient so far left unspecified is the set of
odd-integer values of the $\zeta_m$. A full understanding of their
values probably requires a direct derivation of $\mathcal{L}_{{\rm
deconfined}}$ from the underlying lattice model, as discussed in
Section III.B of Ref.~\cite{psgbosons}, without the long detour
taken here into dual vortex variables. We defer such an analysis to
future work, but will now note some important restrictions on the
values of the $\zeta_m$.

To obtain these restrictions, let us consider the structure of
vortices that can be created from the $\xi_m$ boson fields, using
arguments similar to those presented in Section~\ref{dta}. Because
these boson fields descended from a duality transformation of $q'$
vortex fields $\overline{\varphi}_m$ each carrying physical magnetic
flux $hc/2e$, we expect that the same vortex structure should also
emerge from an analysis of vortex saddle points of
Eq.~(\ref{moose2}). That this is indeed the case was explained in
Ref.~\cite{psgbosons}, and we now review the argument. An elementary
vortex is created by inducing a $2 \pi$ winding in the phase of
$\xi_0$ (say), while keeping the remaining $(q'-1)$ boson fields
$\xi_{m\neq 0}$ topologically trivial. (Moving the phase winding to
the other $(q'-1)$ fields yields a total of $q'$ distinct elementary
vortices, as expected.) Then, under the action in
Eq.~(\ref{moose2}), the $\widetilde{\mathcal{A}}^m_\mu$ gauge fields
will respond to minimize the total action of this vortex state. At a
large distance, $r$, from the vortex center, we expect that these
fields will be pure gauge, and oriented purely in the azimuthal (the
angular co-ordinate $\theta$) direction; keeping in mind the
symmetry of the this vortex state and the constraint in
Eq.~(\ref{gaugeconstraint1}) we obtain
\begin{eqnarray}
\widetilde{\mathcal{A}}^0_\theta (r \rightarrow \infty) &=&
\frac{A}{r} \nonumber \\
\widetilde{\mathcal{A}}^m_\theta (r \rightarrow \infty) &=&
-\frac{A}{(q'-1)r}~~~~, m\neq 0, \label{Alarge}
\end{eqnarray}
where $A$ is a constant to be determined by minimizing the action.
The $1/r$ decay above ensures that $\oint dr_\mu
\widetilde{\mathcal{A}}^m_\mu$ is a constant on a contour far from
the center of the vortex and measures the total
$\widetilde{\mathcal{A}}^m_\mu$ flux trapped near the center of the
vortex. Inserting the above configurations of $\xi_m$ and
$\widetilde{\mathcal{A}}^m_\mu$ into the action in
Eq.~(\ref{moose2}), we find that the total energy of a single vortex
is logarithmically divergent, and the optimal vortex configuration
will minimize the co-efficient of this logarithmic divergence. The
co-efficient is proportional to \cite{psgbosons}
\begin{equation}
(1-A)^2 + \sum_{m=1}^{q'-1} (A/(q'-1))^2 , \label{valA}
\end{equation}
and minimizing this expression yields $A = 1-1/q'$. Upon including
the physical electromagnetic field, the logarithmic divergence is
cutoff by the London penetration depth, and the total magnetic flux
will be that required to saturate the remaining phase winding of the
$\xi_m$ at large $r$; a simple calculation from the result in
Eq.~(\ref{valA}) shows that this yields the required flux of
$hc/2e$.

Turning to our objective of restricting the values of the $\zeta_m$,
let us now insert the results above for the values of
$\widetilde{\mathcal{A}}^m_\mu$ into Eq.~(\ref{gaugeconstraint2}),
and so obtain the value of the total $a_\mu$ flux associated with
the vortex:
\begin{eqnarray}
\oint dr_\mu a_\mu &=& \pi \left( \zeta_0 A + \sum_{m=1}^{q'-1}
\zeta_m \left[ - \frac{A}{(q'-1)} \right] \right) \nonumber \\
&=& \pi \left( \zeta_0 - \frac{1}{q'} \sum_{m=0}^{q'-1} \zeta_m
\right)
\end{eqnarray}
This total $a_\mu$ flux is observed by the $\Psi$ fermions to be
trapped at each vortex. Because the $\Psi$ fermions pick up a -1
Berry phase around each vortex, and recalling the constraint that
all the $\zeta_m$ have to be odd integers, we obtain the additional
constraint that
\begin{equation}
 \frac{1}{q'} \sum_{m'=0}^{q'-1} \zeta_m' = \mbox{even integer}. \label{zetaconstraint}
\end{equation}
The choice $\zeta_m=1$ for all $m$ does not satisfy this constraint.
However, choosing an equal number of $\zeta_m=1$ and $\zeta_m=-1$ so
that $\sum_m \zeta_m = 0$, does satisfy Eq.~(\ref{zetaconstraint});
explicit choices of this type were discussed at the end of
Section~\ref{sec:vbsferm}. It is also clear that no solution is
possible for odd $q'$, indicating that a deconfined critical point
does not exist in this case.

\section{Doping the staggered flux spin liquid}
\label{sf2}

The staggered flux (sF) spin liquid has been considered in some
detail in the context of the ``SU(2) slave particle'' description of
Wen, Lee and collaborators \cite{su2,nagaosa}. We will begin with
the same theory in the undoped insulator, but use a vortex theory to
describe the doped system. As noted earlier, aspects of our analysis
will be uncomplete, because the PSG of the monopoles above the sF
phase is not available. Our analysis will address a variety of
transitions out of a finite doping superconducting state. The
resulting quantum critical points are in a different universality
class from the zero-doping quantum critical points which have been
proposed elsewhere as the fixed points controlling the finite doping
physics \cite{senthillee}.

\subsection{Undoped insulator}
\label{sf3}

The staggered flux spin liquid is described by a mean-field
Hamiltonian expressed in terms of a $S=1/2$ fermionic spinor $f_{j
\sigma}$:
\begin{eqnarray}
  \label{eq:sflux}
  H_{sF} &=& \sum_j \big\{ -t( f_{j \sigma}^\dagger f_{j+\hat{x},\sigma} + f_{j \sigma}^\dagger
  f_{j+\hat{y},\sigma}) \nonumber \\
  &~& -i\eta_j t' (f_{j \sigma}^\dagger f_{j+\hat{x},\sigma} - f_{j \sigma}^\dagger
  f_{j+\hat{y},\sigma}) + {\rm h.c.}\big\}
\end{eqnarray}
For $t'\neq t$, these fermions see a staggered flux ($\neq \pi$)
which apparently breaks translational symmetry.  Since the sign of
this flux cannot be removed by a U(1) gauge transformation,
translational symmetry is implemented instead with a particle/hole
transformation. The PSG transformations which leave $H_{sF}$
invariant are
\begin{eqnarray}
T_x&:& f_{j\sigma} \rightarrow i \eta_{j'} \epsilon_{\sigma \sigma'}
f_{j',\sigma}^\dagger
\nonumber \\
T_y&:& f_{j\sigma} \rightarrow i \eta_{j'} \epsilon_{\sigma
\sigma'} f_{j',\sigma}^\dagger \nonumber \\
R_{\pi/2}^{\rm dual} &:& f_{\sigma} \rightarrow if_{\sigma}\nonumber \\
I_x^{\rm dual} &:& f_{j\sigma} \rightarrow i \eta_{j'}
\epsilon_{\sigma \sigma'} f_{j',\sigma}^\dagger \nonumber \\
\mathcal{T}&:& f_{j\sigma} \rightarrow i \eta_j
f_{j\sigma}~~;~~f_{j\sigma}^\dagger \rightarrow i \eta_j
f_{j\sigma}^\dagger , \label{psgfsf}
\end{eqnarray}
As in Eq.~(\ref{psgf}), all operations are canonical apart from
time-reversal, and for this case the transformation of
$f_{\sigma}^\dagger$ is independent of $f_\sigma$. In all cases, the
site $j'$ is the image of the site $j$ under the noted
transformation {\em e.g.\/} $j^\prime = j + \hat{x}$ under $T_x$.

Moving beyond mean field theory, the $f_\sigma$ fermions are coupled
to a U(1) gauge field $\mathcal{A}_\mu$, under which they have a
uniform charge of +1 (see Table~\ref{table1}), in contrast to the
charge of $\eta_j$ obtained in the VBS case. The PSG transformation
of $\mathcal{A}_\mu$ are easily deduced from those of the $f_\sigma$
by imposing the requirement of gauge invariance. The completion of
the PSG Eq.~(\ref{ta2}), as in Eq.~(\ref{psgcalA}), is
\begin{eqnarray}
T_x&:& \mathcal{A}_\mu \rightarrow - \mathcal{A}_\mu \nonumber \\
T_y&:& \mathcal{A}_\mu \rightarrow - \mathcal{A}_\mu \nonumber \\
R_{\pi/2}^{\rm dual}&:&\mathcal{A}_\tau\rightarrow
\mathcal{A}_\tau~;~\mathcal{A}_x\rightarrow \mathcal{A}_y~;~
\mathcal{A}_y\rightarrow -\mathcal{A}_x
\nonumber \\
I_{x}^{\rm
dual}&:&\mathcal{A}_\tau\rightarrow-\mathcal{A}_\tau~;~\mathcal{A}_x\rightarrow\mathcal{A}_x~;~
\mathcal{A}_y\rightarrow-\mathcal{A}_y
\nonumber \\
\mathcal{T}&:&\mathcal{A}_\tau \rightarrow
-\mathcal{A}_\tau~;~\mathcal{A}_x \rightarrow
\mathcal{A}_x~;~\mathcal{A}_y \rightarrow \mathcal{A}_y.
\label{psgcalAs}
\end{eqnarray}

Finally, to complete the analysis of the sF spin liquid, we need the
transformations of the monopoles in $\mathcal{A}_\mu$, the analog of
the relations in Eq.~(\ref{psgm}). These we do not present here, but
our analysis below can easily be extended to include them.

\subsection{Doped staggered flux state}
\label{sec:dope2}

As stated earlier, we dope the spin liquid state by introducing two
species of holon bosons, $b_{1,2}$, which carry charges $\pm 1$
under the $\mathcal{A}_\mu$ gauge field. The PSG of these bosons can
be deduced by the requirement that the physical electron operator be
invariant under all transformations. The latter is connected to the
slave particles by \cite{su2,psgdimers}
\begin{equation}
  \label{eq:cdef}
  c_{j\sigma}  =  b_{j1}^\dagger f_{j\sigma} + i \eta_j b_{j2}^\dagger
  \epsilon_{\sigma\sigma'} f_{j\sigma'}^\dagger,
\end{equation}
$ $From Eqs. (\ref{psgfsf}) and (\ref{eq:cdef}) we can deduce
\begin{eqnarray}
T_x&:& b_1 \rightarrow b_2~;~b_2 \rightarrow b_1 \nonumber \\
T_y&:& b_1 \rightarrow b_2~;~b_2 \rightarrow b_1  \nonumber \\
R_{\pi/2}^{\rm dual}&:&b_1 \rightarrow b_1~;~b_2 \rightarrow b_2
\nonumber \\
I_{x}^{\rm dual}&:& b_1 \rightarrow b_2~;~b_2 \rightarrow b_1
\nonumber \\
\mathcal{T}&:&b_1 \rightarrow b_2^\dagger~;~b_2 \rightarrow
b_1^\dagger . \label{psgbs}
\end{eqnarray}
Note that the differences from the corresponding Eq.~(\ref{psgb})
for the VBS case are restricted to $R_{\pi/2}^{\rm dual}$ and
$\mathcal{T}$. In Eq.~(\ref{psgb}), the $1,2$ vortex flavors are
interchanged under $R_{\pi/2}^{\rm dual}$ but not under
$\mathcal{T}$, while in Eq.~(\ref{psgbs}) above the opposite is
true.

Proceeding with the duality from bosons $b_{1,2}$ to vortices
$\psi_{1,2}$ and U(1) gauge fields $A_{1\mu}$ and $A_{2\mu}$ as in
Section~\ref{sec:dope1}, the PSG of the vortices in
Eq.~(\ref{psgpsi}) now becomes
\begin{eqnarray}
T_x&:& \psi_1 \rightarrow \psi_2~;~\psi_2 \rightarrow \psi_1 \nonumber \\
T_y&:& \psi_1 \rightarrow \psi_2~;~\psi_2 \rightarrow \psi_1  \nonumber \\
R_{\pi/2}^{\rm dual}&:&\psi_1 \rightarrow \psi_1~;~\psi_2
\rightarrow \psi_2
\nonumber \\
I_{x}^{\rm dual}&:& \psi_1 \rightarrow \psi_2^\dagger~;~\psi_2
\rightarrow \psi_1^\dagger
\nonumber \\
\mathcal{T}&:&\psi_1 \rightarrow \psi_2~;~\psi_2 \rightarrow \psi_1,
\label{psgpsis}
\end{eqnarray}
while the PSG of the dual U(1) gauge fields $A_{1,2}$ in
Eq.~(\ref{psgA}) is replaced here by
\begin{eqnarray}
T_x&:& A_{1\mu} \rightarrow A_{2\mu}~;~A_{2\mu} \rightarrow A_{1\mu} \nonumber \\
T_x&:& A_{1\mu} \rightarrow A_{2\mu}~;~A_{2\mu} \rightarrow A_{1\mu} \nonumber \\
R_{\pi/2}^{\rm dual}&:& A_{1\tau} \rightarrow A_{1\tau}~;~A_{1x}
\rightarrow
A_{1y}~;~A_{1y} \rightarrow -A_{1x}~;~\nonumber \\
&~& A_{2\tau} \rightarrow A_{2\tau}~;~A_{2x} \rightarrow
A_{2y}~;~A_{2y} \rightarrow -A_{2x}
\nonumber \\
I_{x}^{\rm dual}&:& A_{1\tau} \rightarrow -A_{2\tau}~;~A_{1x}
\rightarrow
A_{2x}~;~A_{1y} \rightarrow -A_{2y}~;~\nonumber \\
&~& A_{2\tau} \rightarrow -A_{1\tau}~;~A_{2x} \rightarrow
A_{1x}~;~A_{2y} \rightarrow -A_{1y}
\nonumber \\
\mathcal{T}&:& A_{1\tau} \rightarrow -A_{2\tau}~;~A_{1x} \rightarrow
A_{2x}~;~A_{1y} \rightarrow A_{2y}~;~\nonumber \\
&~& A_{2\tau} \rightarrow -A_{1\tau}~;~A_{2x} \rightarrow
A_{1x}~;~A_{2y} \rightarrow A_{1y}. \label{psgAs}
\end{eqnarray}
Again, note that the differences from Eqs.~(\ref{psgpsi},\ref{psgA})
for the VBS case are restricted to $R_{\pi/2}^{\rm dual}$ and
$\mathcal{T}$. In Eqs.~(\ref{psgpsi},\ref{psgA}), the $1,2$ vortex
flavors are interchanged under $R_{\pi/2}^{\rm dual}$ but not under
$\mathcal{T}$, while in Eqs.~(\ref{psgpsis},\ref{psgAs}) above the
opposite is true.

Just as in Section~\ref{sec:dope1}, we now account for the influence
of the mean hole density of $\delta$ (with $\delta/2 = p/q$) on the
vortices by introducing $2q$ flavors of vortices $\varphi_{1\ell}$
and $\varphi_{2\ell}$. The PSG analog of the transformations of
these vortices in Eq.~(\ref{psgvarphi}) is now
\begin{eqnarray}
T_x&:& \varphi_{1\ell} \rightarrow \varphi_{2,\ell+1}~;~
\varphi_{2\ell} \rightarrow \varphi_{1,\ell+1} \nonumber \\
T_y&:& \varphi_{1\ell} \rightarrow \varphi_{2\ell} \omega^{-\ell}
~;~\varphi_{2\ell} \rightarrow \varphi_{1\ell} \omega^{-\ell}  \nonumber \\
R_{\pi/2}^{\rm dual}&:&\varphi_{1\ell} \rightarrow
\frac{1}{\sqrt{q}} \sum_{m=0}^{q-1} \varphi_{1m} \omega^{-\ell
m}~;~\nonumber \\
&~& \varphi_{2\ell} \rightarrow \frac{1}{\sqrt{q}} \sum_{m=0}^{q-1}
\varphi_{2m} \omega^{-\ell m}
\nonumber \\
I_{x}^{\rm dual}&:& \varphi_{1\ell} \rightarrow
\varphi_{2\ell}^\dagger~;~\varphi_{2\ell} \rightarrow
\varphi_{1\ell}^\dagger
\nonumber \\
\mathcal{T}&:&\varphi_{1\ell} \rightarrow
\varphi_{2\ell}~;~\varphi_{2\ell} \rightarrow \varphi_{1\ell},
\label{psgvarphis}
\end{eqnarray}
Once again, the differences from Eq.~(\ref{psgvarphi}) for the VBS
case are the opposite treatment of the $1,2$ vortex flavors between
$R_{\pi/2}^{\rm dual}$ and $\mathcal{T}$.

\subsection{Low energy theory}
\label{sec:lw2}

Following the analysis in Section~\ref{sec:lw1}, we now need to
write down the most general effective action consistent with the PSG
and gauge charge assignments obtained in Sections~\ref{sf3} and
\ref{sec:dope2}. This action can then describe transitions out of a
$d$-wave superconductor into supersolid and insulating phases
affiliated with the staggered flux phase.

The required action continues to have the form in
Eq.~(\ref{vbsaction}), with the contributions $\mathcal{L}_{\rm
U(1)}$ and $\mathcal{L}_{2 \varphi}$ retaining their forms in Eqs.
(\ref{la}) and (\ref{l2a}). The quartic term $\mathcal{L}_{4
\varphi}$ will remain unspecified as the most general quartic
polynomial in $\varphi$ which is invariant under
Eq.~(\ref{psgvarphis}).

The fermionic contribution $\mathcal{L}_f [\mathcal{A}_\mu ]$ is
obtained by taking the continuum limit of Eq. (\ref{eq:sflux}) and
yields the familiar Dirac form at the four nodal points, which was
schematically indicated in Eq.~(\ref{sdirac}).

The additional quadratic invariant, $\widetilde{\mathcal{L}}_{2
\varphi}$, in Eq.~(\ref{l2a}) is now no longer given by
Eq.~(\ref{l2p}). Instead, now there is a term whose origin is the
staggered flux acting on the holons. Upon dualizing to vortices,
this flux becomes equivalent to a staggered ``chemical potential''
acting on the vortices. A search for such terms reveals the
following contribution which is present {\em only \/} when $q$ is a
multiple of 4:
\begin{eqnarray}
&& \widetilde{\mathcal{L}}_{2\varphi} = h_s \sum_{\ell=0}^{q-1}
(-1)^\ell \Biggl[ \varphi_{1,\ell + q/2}^{\ast} \left(\frac{\partial
}{\partial \tau} - i B_{s\tau}- i B_{a\tau} \right) \varphi_{1\ell}
\nonumber
\\ &&~~~~~~~~~~~~- \varphi_{2,\ell + q/2}^{\ast} \left(\frac{\partial }{\partial
\tau}- i B_{s\tau} + i B_{a\tau} \right) \varphi_{2\ell} \Biggr].
\label{ll2}
\end{eqnarray}
The PSG transformation associated with time-reversal allows only
Eq.~(\ref{ll2}) in the present case, and only Eq.~(\ref{l2p}) in the
doped VBS case.

The final ingredient are the monopole terms $\mathcal{L}_m$ which
couple monopoles to vortex bilinears. As in the VBS case, the vortex
bilinears must carry dual staggered gauge ($B_{a\mu}$) charge $\pm
2$, and so must mix the 1,2 vortex types. We will leave such terms
undetermined here, but expect their influence can be easily included
in the analysis below.

\subsection{Vortex spectrum and quantum phase transitions out of the
superconductor} \label{vortexs}

As in the VBS case, it is useful to divide the discussion into
various classes of values of $q \mbox{(mod 4)}$.

\subsubsection{$q =1,3~\mbox{(mod 4)}$}
\label{sec:sfqodd}

Although we do not have the explicit form of the monopole terms,
$\mathcal{L}_m$, available, it appears a safe assumption that such
terms will not contribute for $q=1,3~\mbox{(mod 4)}$. The reason for
this is similar to that for the VBS case: the wavevectors associated
with the vortex fields $\varphi_{1,2\ell}$ are not expected to match
those of the monopole transformations.

The `staggered flux' term in Eq.~(\ref{ll2}) also does not
contribute. So there are $2q$ vortex species, just as in the VBS
case. The low energy theory is given by the sum of Eqs. (\ref{la}),
(\ref{l2a}), (\ref{sdirac}) and $\mathcal{L}_{4 \varphi}$. This
theory will describe transitions from the $d$-wave superconductor
into proximate insulating or supersolid phases.

The present theory is closely related to the critical theory in
Eq.~(\ref{dSCVBS}) for the doped VBS case. Here the mutual
statistics between the vortices and the Dirac fermions is
implemented by $B_{a\mu}$ gauge field, and $\varphi_1$ and
$\varphi_2$ vortices have opposite charges under $B_{a \mu}$. In
contrast, the mutual statistics in Eq.~(\ref{dSCVBS}) is implemented
by $\alpha_\mu$, and the vortices had charges $\zeta_m$. So the
present case corresponds to choosing $\zeta_m=1$ for the $\varphi_1$
vortices and $\zeta_m=-1$ for the $\varphi_2$ vortices. The only
remaining difference between the theories then are the differences
between the PSGs of the vortices in Eqs. (\ref{psgvarphi}) and
(\ref{psgvarphis}). This will lead to minor differences in the range
of the competing orders which can appear in the insulating phases.

While the above derivation of the critical theory of fermions and
vortices does have the advantage of preserving the lattice PSG at
all stages, it does have the unphysical feature that the nodal
points are pinned at the wavevectors $(\pm \pi/2, \pm \pi/2)$ even
in the finite doped superconducting case. A different continuum
limit, along the lines of that discussed in
Section~\ref{sec:vbsferm} is needed to rectify this defect.

We can undualize the critical theory to obtained the theory for the
deconfined critical point as in Eq.~(\ref{moose2}): the only changes
are that the values of $\zeta_m$ are as specified above, the $a_\mu$
gauge field is replaced by $\mathcal{A}_\mu$ with $\mathcal{A}_\mu$
obeying the constraint Eq.~(\ref{gaugeconstraint2}), and the fermion
term $\mathcal{L}_c[a_\mu]$ takes the Dirac from $\mathcal{L}_\Psi [
\mathcal{A}_\mu ]$ in Eq.~(\ref{sdirac}).

\subsubsection{$q = 2~\mbox{(mod 4)} $}
\label{sec:sfq2}

The `staggered flux' term in Eq.~(\ref{ll2})  does not contribute
for this case either. However, we expect non-trivial contributions
from the monopole terms in $\mathcal{L}_m$ now. It is a plausible
hypothesis that such terms will reduce the degeneracy of the vortex
spectrum from $2q$ to $q/2$, as was the case for the VBS state,
reviewed in Table~\ref{table2}.

\subsubsection{$q = 0~\mbox{(mod 4)} $}
\label{sec:sfq4}

This is the case most relevant for application to the cuprates, and
displays some rather interesting features, not encountered in any of
the cases considered so far.

Now we have to include the `staggered flux' term in Eq.~(\ref{ll2}).
This term has first order time derivatives, and so raises the
possibility that the vortices will have a `non-relativistic'
dispersion spectrum.

We also have to consider the possible influence of monopole terms,
$\mathcal{L}_m$ here. However, we will see below that the `staggered
flux' term in Eq.~(\ref{ll2}) is already sufficient to reduce the
degeneracy of the vortex species from $2q$ to $q$, as required by
the degeneracy spectrum for this value of $q$ from
Table~\ref{table2}. Therefore, it is not an unreasonable expectation
that the monopole terms will not contribute here. In any case, one
can view the following analysis as a diagonalization of the existing
terms in the vortex action, which can be useful basis for
considering the subsequent possible influence of monopole terms.

It is useful to work with the following parameterization to
diagonalize the vortex quadratic form in Eqs.~(\ref{l2a}) and
(\ref{ll2}). We work with the four complex fields $W_m$, $X_m$,
$Y_m$, and $Z_m$, with $m = 0 \ldots q/2-1$, defined by
\begin{eqnarray}
\varphi_{1m} &=& \frac{W_m + X_m^{\ast}}{\sqrt{2 h_s}} \nonumber \\
\varphi_{1,m+q/2} &=& (-1)^m \frac{W_m - X_m^{\ast}}{\sqrt{2 h_s}} \nonumber \\
\varphi_{2m} &=& \frac{Y_m + Z_m^\ast}{\sqrt{2 h_s}} \nonumber \\
\varphi_{2,m+q/2} &=& - (-1)^m \frac{Y_m - Z_m^\ast}{\sqrt{2 h_s}}
\label{e12}
\end{eqnarray}
Upon inserting this parameterization into Eqs.~(\ref{l2a}) and
(\ref{ll2}), and ignoring second-order time derivative terms which
are unimportant at low energies, the quadratic vortex Lagrangian
becomes
\begin{eqnarray} && \mathcal{L}_{2 \varphi} =
 \sum_{m=0}^{q/2-1} \Biggl[ \label{vortexnr} \\
&& ~ W_m^{\ast}\left(\frac{\partial }{\partial \tau} - i B_{s\tau}-
i B_{a\tau} \right) W_m + \frac{1}{h_s} \left| (
\partial_i - i B_{si} - i B_{ai})
W_m \right|^2 + \frac{s}{h_s} |W_m|^2 \nonumber \\
&& + X_m^{\ast}\left(\frac{\partial }{\partial \tau} + i B_{s\tau} +
i B_{a\tau} \right) X_m + \frac{1}{h_s} \left| ( \partial_i + i
B_{si} + i B_{ai})
X_m \right|^2 + \frac{s}{h_s} |X_m|^2 \nonumber \\
&& + Y_m^{\ast}\left(\frac{\partial }{\partial \tau} - i B_{s\tau} +
i B_{a\tau} \right) Y_m + \frac{1}{h_s} \left| ( \partial_i - i
B_{si} + i B_{ai})
Y_m \right|^2 + \frac{s}{h_s} |Y_m|^2 \nonumber \\
&& + Z_m^{\ast}\left(\frac{\partial }{\partial \tau} + i B_{s\tau} -
i B_{a\tau} \right) Z_m + \frac{1}{h_s} \left| ( \partial_i + i
B_{si} - i B_{ai}) Z_m \right|^2 + \frac{s}{h_s} |Z_m|^2 \Biggr]
\nonumber
\end{eqnarray}
The full low-energy theory is now the sum of Eqs. (\ref{la}),
(\ref{sdirac}), (\ref{vortexnr}) and $\mathcal{L}_{4 \varphi}$.

We note from Eq. (\ref{vortexnr}) that the fields $X_m$, $W_m$,
$Y_m$, and $Z_m$ are canonical non-relativistic Bose fields. From
their $B_{s\mu}$ charges, we deduce that the $W_m$ and $Y_m$ bosons
are vortices, while the $X_m$ and $Z_m$ bosons are anti-vortices.
There are, therefore, a total of $q$ flavors of vortices and
anti-vortices. The total count of degenerate vortex/anti-vortex
excitations is therefore the same as that obtained for the VBS case,
where we also had $q'=q$ for the case where $q$ was a multiple of 4.

However, the non-relativistic nature of Eq.~(\ref{vortexnr}) has
important and novel consequences for nature of the fluctuations in
the superconductor. For $s>0$, the ground state of
Eq.~(\ref{vortexnr}) is exactly the vacuum of the $W_m$, $X_m$,
$Y_m$, and $Z_m$ bosons. Consequently, virtual quantum fluctuations
of low energy vortex-anti-vortex pairs are essentially totally
absent in the superconductor---this is in strong contrast to all
other cases with a `relativistic' action, where such fluctations
dominate and drive the superconductor-insulator transition.

In the absence of such vacuum fluctuations, the gauge-field
interactions between the vortices and anti-vortices are unscreened,
and so will lead to the formation of vortex-anti-vortex bound states
which will remain robust as the value of $s$ is lowered. Indeed, a
glance at Eq.~(\ref{vortexnr}) shows that we can expect that the
lowest energy bound states to form between the $W_m$ and the $X_m$
bosons and between the $Y_m$ and the $Z_m$ bosons. With the lowering
of $s$, it is the energy of these bound states which will first
cross zero. As the net vorticity of this condensing boson is zero,
the this is a transition from the superconductor to a {\em
supersolid}. Thus we have reached the remarkable conclusion that the
transition out of the sF-doped $d$-wave superconductor is
necessarily into a supersolid, for the case that $q$ is a multiple
of 4.

We now ask whether the above spectra of low energy vortices and
anti-vortices places any restrictions on the nature of density wave
order in the supersolid. As in Ref.~\cite{psgbosons} we can define
the density wave order operators $\rho_{mn}$, with $m,n = 0,1,
\ldots ,q-1$, at the wavevectors $Q_{mn} = (2 \pi p/q) (m,n)$ by
\begin{equation}
\rho_{mn}= \omega^{mn/2} \sum_{\ell = 0}^{q-1} \left[
\varphi^{\ast}_{1 \ell} \varphi_{1,\ell+n} + \varphi^{\ast}_{2 \ell}
\varphi_{2,\ell+n} \right] \omega^{\ell m}. \label{e11}
\end{equation}
These operators follow from the requirement that the PSG
transformations in Eq.~(\ref{psgvarphis}) lead to transformations
for $\rho_{mn}$ required for conventional density wave operators
\cite{psgbosons}. We now need to insert the parameterization
(\ref{e12}) into (\ref{e11}), and thence deduce the nature of
$\rho_{mn}$ fluctuations under Eq.~(\ref{vortexnr}). The expressions
so obtained are quite lengthy, but we can understand the general
result by considering a few representative cases. So for {\em e.g.}
$q=4$, we find
\begin{eqnarray}
\rho_{11} &=& e^{i \pi/4} \left( W_0^\ast  W_1 + i W_1^\ast W_0  +
X_1^\ast X_0 -i X_0^\ast X_1  \right) + \left( W \rightarrow Y, X
\rightarrow Z \right), \label{rho11}
\end{eqnarray}
while
\begin{eqnarray}
\rho_{10} &=& W_0 X_0 + i W_1 X_1 + W_0^\ast X_0^\ast + i W_1^\ast
X_1^\ast + \left( W \rightarrow Y, X \rightarrow Z \right).
\label{rho10}
\end{eqnarray}
Notice, a crucial difference between the two cases considered above.
In Eq.~(\ref{rho11}) we only have combinations between $W$, $X$,
$Y$, $Z$ creation and annihilation operators; however, it is
impossible to annihilate such bosons from the superconductor vacuum,
and so can expect that $\rho_{11}$ fluctuations are strongly
suppressed. In contrast, $\rho_{10}$ involves combinations of
operators which are always both creation or annihilation operators;
in particular there are terms which lead to the creation of the $W$,
$X$ and $Y$, $Z$ low energy bound states that were noted above. We
therefore conclude that $\rho_{10}$ fluctuations are strongly
enhanced as $s$ is lowered.

These observations can be extended into a simple general result. Let
us define the susceptibility $\chi_{mn}$ as the correlator of
$\rho_{mn}$ and $\rho_{mn}^\ast = \rho_{-m,-n}$ at zero external
frequency and momentum. Then, evaluating one-loop $\chi_{mn}$ under
the Lagrangian in Eq.~(\ref{vortexnr}) we find
\begin{equation}
\chi_{mn} = 0, ~\mbox{for $m+n$ even},
\end{equation}
while there is a divergent response at other values of $m$, $n$ as
$s \rightarrow 0$:
\begin{eqnarray}
\chi_{mn} &\sim& \int \frac{d \Omega d^2 k}{8 \pi^3}
\frac{1}{(\Omega^2 + (k^2 + s)^2/h_s^2)} \nonumber \\
&\sim& \ln (\Lambda/s), ~~~~\mbox{for $m+n$ odd},
\end{eqnarray}
with $\Lambda$ an upper cutoff. Unless the above effects are
overwhelmed by anomalously large quartic couplings in
$\mathcal{L}_{4 \varphi}$, we conclude that the transition from the
doped sF $d$-wave superconductor is into a supersolid in which the
strongest density modulations are at wavevectors $Q_{mn}$ with $m+n$
odd.

We reiterate that while it is plausible that the monopoles do not
modify the above conclusions, this has not yet been firmly
established.

\section{Conclusions}
\label{conc}

This paper has presented a new approach to the physics of doped U(1)
spin liquids.

We began with the two most popular examples of U(1) spin liquids on the
square lattice at a density of one electron per site. The first, dubbed
the VBS state, is expressed as a pure compact U(1) gauge theory;
condensation of monopoles leads to confinement of spinons and the
appearance of Valence-Bond-Solid (VBS) order at low energies. The
second, the staggered flux (sF) state, has 4 species of gapless Dirac
fermion spinon excitations. It is possible that these fermions suppress
monopole condensation in the spin liquid, so that there is no
confinement, fractionalized gapless excitations survive, and there is no
broken lattice symmetry.  Even if the sF state is unstable, it still
forms the basis for a quantum critical point, upon which to base an
effective field theory in the spirit of this paper.  In any case, it is
only {\sl more} stable than the VBS liquid (which definitely is unstable
at low energies to long-range VBS order).

Essential characteristics of the VBS and sF states are their
transformation properties (the `PSG') of the monopole tunnelling
events under the symmetries of the square lattice. The existence of
a non-trivial PSG for the monopole led to the appearance of VBS
order in phases where the monopoles condense.

Then we doped the spin liquids with holes of rational density
$\delta$, obeying Eq.~(\ref{defpq}), which defines the crucial
integer-valued parameter, $q$. The hole degrees of freedom are most
naturally expressed in terms of charge $e$, spinless bosons
(`holons'), although this does not imply existence of quasiparticles
with these quantum numbers in any phase. The bosons appear in two
species, each with density $\delta/2$, carrying opposite gauge
charges under the U(1) gauge field of the spin liquid. After a
duality transformation, the charge degrees of freedom were
encapsulated in the dynamics of vortices, each carrying magnetic
flux $hc/(2e)$. Central to our analysis were the PSG transformations
of these vortices, and the couplings between monopoles and vortices
that were allowed by the PSG.

By an analysis of the low energy theories allowed by the PSG of the
monopoles, vortices, and fermionic spinons, we arrived two main
classes of results, which are summarized in the subsections below.

\subsection{Vortex spectrum in the superconductor}
\label{conc1}

A recurring approach in condensed matter physics is the formulation of
effective models in terms of ``elementary'' excitations.  In some cases
(e.g. Fermi liquids), such excitations are adiabatically connected to
free electrons.  Other elementary excitations are particular to specific
states of matter, e.g. collective Goldstone modes resulting from
symmetry breaking, or topological excitations such as domain walls in
one dimension.  The set of elementary excitations is usually
considered to be a fundamental characteristic of a particular phase of
matter.  Most theories of quantum critical points begin by
assuming the gap of some set of these excitations is tuned
parametrically to zero, and the fields of the critical theory are
in correspondence with them.

Whenever gapless excitations are present, there is potential for
significant ambiguity in the identification of elementary excitations,
because new ``particles'' can be built from collections of a large
number of very low energy ones, to form a new basis.  This occurs
prominently in the theory of one-dimensional systems, where various
forms of {\sl bosonization} trade gapless fermionic and bosonic field
variables.  In this paper, we sought to obtain a description of the
elementary excitations of two dimensional superconductors on the square
lattice.   Clearly this must include both electronic quasiparticles and
the vortices.  When the former are gapless, we again have potential
difficulties of the type described above.  This paper presents one
approach to deal with this formidable problem.

The approach builds upon a previous paper, \cite{psgbosons} in which
vortices of super{\sl fluid} phases of {\sl boson} models on the square
lattice were classified.  A central result was that bosons of density
$p'/q'$ ($p'$, $q'$ relatively prime integers) led to vortices with a
$q'$-fold degeneracy. Unitary transformations within this vortex
`flavor' space encoded various space group operators. Furthermore, a
vortex with a fixed orientation in flavor space necessarily had static
modulations at wavevectors $(2 \pi p'/q')(m,n)$ (with $m$, $n$,
integers) in all spin-singlet observables in its vicinity. These
modulations can be viewed as a strong coupling analog of Friedel
oscillations around impurities in Fermi liquids.  Quantum criticality
from superfluid to insulating states can be formulated in the $q$-vortex
variables.

A ``direct'' attack on the analogous problem in a superconductor would
be to try to regard the Cooper pair as the boson of
Ref.~\cite{psgbosons}.  This has the difficulty that the Cooper pair
field is coupled in a very strong fashion (as in the Bogoliubov-de
Gennes equations) to the quasiparticles.  The attempt to extricate the
collective vortex excitations from the quasiparticles in this
approach is quite non-trivial.  Instead, we chose to break down the
problem in two steps.  First, we reformulated the microscopic electronic
model using gauge theoretical methods to {\sl fractionalize} the
electron into a bosonic charge $e$ holon and a neutral spin-$1/2$
spinon.  A superconducting state is then obtained as a superfluid state
of the holons.  This has the advantage that the spinons, which become
the quasiparticles in the superconducting state, are coupled {\sl
  indirectly} to the holons through a gauge field.  It is then possible
to treat the holons themselves by duality techniques directly analogous
to those of Ref.~\cite{psgbosons}.

An interesting wrinkle in this procedure is that many different choices
for the initial fractionalization procedure are possible.  Although all
of these are microscopically equivalent, saddle point and other
approximations are inevitably required, leading different choices to
more naturally describe different {\sl insulating} states.  Thus
dependent upon the sort of Mott insulator reached from the
superconductor, a particular form of fractionalization may be most
appropriate.  In the superfluid phase obtained by condensing the holons,
however, we believe the same {\sl superconducting} phase can be reached
from different fractionalized variables.  Therefore it would be natural
to expect the same vortex excitations in the superconductor to be
obtained from each of these choices.   Despite the simplicity of this
expectation, its correctness is by no means transparent in the actual
calculations. Nevertheless, to the extent we are able to check this, we
do indeed find agreement between the different gauge decouplings.

For the case of a superconductor obtained by doping a VBS state, the
intermediate steps of our analysis were quite involved, but our
final result was simple. The degeneracy and PSG of the low energy
vortex excitations were identical to that in pure boson models, with
a density of bosons equal to the density of electron pairs {\em
i.e.\/} the integer $q'$ is determined here in
Eq.~(\ref{defqprime}). Consequently many of the results of
Ref.~\cite{psgbosons} can be applied to this electronic model
without modification.

The results for the doped sF state were tentative, pending
determination of the monopole terms. However, for the cases $q=0,1,3
~\mbox{(mod 4)}$ we already obtained a vortex degeneracy identical
to the VBS case in Table~\ref{table2}, plausibly suggesting that
monopole terms can be ignored for these cases.  It seems likely that a
proper treatment of the monopoles would give complete correspondence
with the VBS case, but this is left unresolved here.

\subsection{Quantum phase transitions out of the superconductor}

With the vortex spectrum in hand, we were able to address quantum
phase transitions associated with the condensation of vortices
(leading to a transition from the superconductor directly to an
insulator) or of vortex-anti-vortex pairs (leading to transition to
a supersolid). The critical theory for such a transition depends
upon the fate of the fermionic $S=1/2$ Bogoliubov quasiparticle
excitations of the superconductor. In weak-coupling BCS theory, a
$d$-wave superconductor has gapless, nodal excitations at four
points in the Brillouin zone. If these survive as gapless
excitations all the way up to the quantum critical point of
interest, then their influence has to be considered in the critical
theory, and they could change the universality class. Alternatively,
it is possible that the fermionic excitations are gapped at the
quantum critical point: in this case, they can be safely integrated
out and can be considered irrelevant to the critical theory.

We now list the various quantum critical points obtained in our
analysis in the following subsections. We will subdivide the
discussion under headers indicating the cases they apply to. We will
explicitly write down the leading quadratic terms in the critical
theory for each case in these subsections. Higher order couplings
are also important, and are strongly constrained by the PSG: for
these we refer the reader back to the body of the paper.

\subsubsection{Superconductor-insulator transition, no gapless
fermions} \label{sec:sinf}

\paragraph*{Doped VBS, all $q$:}  As parameters are changed
in a $d$-wave superconductor, it is possible that the electron
pairing becomes short ranged, and the fermionic excitation spectrum
is fully gapped. Such a scenario also appears plausible from the
perspective of a confining VBS insulator that is fully gapped, as it
is moving towards a quantum phase transition into a superconductor.
In this case, the theory for the superconductor-insulator quantum
critical point was found to be identical to that discussed at some
length in Ref.~\cite{psgbosons} for boson models. Note that the
notation in Ref.~\cite{psgbosons} is different from ours here---the
integer parameter $q$ of Ref.~\cite{psgbosons} should be set equal
to the parameter $q'$ defined here in Eq.~(\ref{defqprime}), which
is related to the density of electronic Cooper pairs. Such a quantum
critical point can occur at all values of $q$ (defined here in
Eq.~(\ref{defpq})) for the doped VBS spin liquid. The dual vortex
theory is expressed in terms of $q'$ complex scalars
$\overline{\varphi}_m$ and a U(1) gauge field $B_{s\mu}$:
\begin{eqnarray}
\mathcal{L}_1 &=& \frac{K_s}{2} \left( \epsilon_{\mu\nu\lambda}
\partial_\nu B_{s \lambda} \right)^2 +
\sum_{m=0}^{q'-1} \Bigl\{ \left|\left(\partial_\mu - i B_{s \mu}
\right)\overline{\varphi}_{m}\right|^2 +  s
|\overline{\varphi}_{m}|^2 \Bigr\} \label{l1}
\end{eqnarray}
Here, and in all actions below, the action for the $B_{s\mu}$ field
has been written in a schematic relativistic form, which is
appropriate for short-range interactions between the bosons---the
Coulomb interactions lead to modifications presented in
Ref.~\cite{predrag}. The superconductor-insulator transition is
accessed by tuning $s$.

As discussed in Ref.~\cite{psgbosons}, for certain values of $q'$
which allow a ``permutative'' PSG, the critical point can be of the
`deconfined' variety \cite{senthil1}: in this case it has a direct
formulations in terms of $q'$ complex scalars, $\xi_m$, each
carrying electrical charge $2e/q'$, and coupled to $q'-1$
non-compact U(1) gauge fields, $\widetilde{A}^m_\mu$. The action was
given in Eq.~(\ref{moose2}), but the gapped fermionic terms can be
dropped:
\begin{eqnarray}
\mathcal{L}_{2} &=& \sum_{m=0}^{q'-1} \left[\left| \left(
\partial_{\mu} - i \widetilde{\mathcal{A}}^{m}_{\mu}  \right)
\xi_{m} \right|^2 + \widetilde{s} |\xi_m |^2 \right]  \nonumber \\
&~&~~+ \sum_{m,n=0}^{q'-1} K_{mn} \left( \epsilon_{\mu\nu\lambda}
\partial_{\nu} \widetilde{\mathcal{A}}^{m}_{\lambda} \right)  \left( \epsilon_{\mu\rho\sigma}
\partial_{\rho} \widetilde{\mathcal{A}}^{n}_{\sigma} \right), \label{l2}
\end{eqnarray}
where the gauge fields obey the constraint in
Eq.~(\ref{gaugeconstraint1})
\begin{equation}
\sum_{m=0}^{q'-1} \widetilde{\mathcal{A}}^m_\mu = 0. \label{lc2}
\end{equation}

\subsubsection{Superconductor-insulator transition with gapless
nodal fermions} \label{sec:sif}

\paragraph*{Doped sF, odd $q$:}
The simplest case where the nodal fermions survive all the way to a
quantum critical point to an insulator was for the doped sF spin
liquid, with $q$ odd (discussed in Section~\ref{sec:sfqodd}). In
this case there is no non-vanishing tri-linear coupling between the
monopoles and the vortices, and the derivation of the effective
action is straightforward and preserves the lattice PSG at all
stages. However, such a derivation has the disadvantage that the
nodal points remained pinned at $(\pm \pi/2, \pm \pi/2)$ even in the
finite doping superconductor. An analysis along the lines of
Section~\ref{sec:vbsferm} seems necessary to rectify this defect.

There are a total of $2q$ vortex fields $\varphi_{1\ell}$ and
$\varphi_{2\ell}$ coupled to 2 U(1) gauge fields, $B_{s\mu}$ and
$B_{a\mu}$, and 4 Dirac fermions, $\Psi$, coupled to a U(1) gauge
field $\mathcal{A}_\mu$. The critical action is
\begin{eqnarray}
\mathcal{L}_3 &=& \frac{K}{2} \left( \epsilon_{\mu\nu\lambda}
\partial_\nu \mathcal{A}_\lambda \right)^2 + \frac{K_s}{2} \left( \epsilon_{\mu\nu\lambda}
\partial_\nu B_{s \lambda} \right)^2 + \frac{K_a}{2} \left( \epsilon_{\mu\nu\lambda}
\partial_\nu B_{a \lambda} \right)^2 \nonumber \\ &~&~+ \frac{i}{\pi}
\epsilon_{\mu\nu\lambda} \partial_\mu B_{a \nu}
\mathcal{A}_{\lambda} + \sum_{\ell=0}^{q-1} \Bigl\{
\left|\left(\partial_\mu - i
B_{s \mu} - i B_{a \mu} \right)\varphi_{1\ell}\right|^2 \nonumber \\
&~&~+ \left|\left(\partial_\mu - i B_{s \mu} + i B_{a \mu}
\right)\varphi_{2\ell}\right|^2  + s \left( |\varphi_{1\ell}|^2 +
|\varphi_{2\ell}|^2 \right) \Bigr\} \nonumber
\\
 &-& i \overline{\Psi}
\gamma_\mu \left(
\partial_\mu + i\mathcal{A}_\mu \right) \Psi
\label{l3}
\end{eqnarray}

As above, under suitable conditions requiring the existence of a
permutative PSG, the critical theory of Eq.~(\ref{l3}) can be
undualized into a theory of $2q$ complex scalars $\xi_m$, each
carrying electromagnetic charge $e/q$. The undualized theory is
\begin{eqnarray}
\mathcal{L}_{4} &=& \sum_{m=0}^{2q-1} \left[\left| \left(
\partial_{\mu} - i \widetilde{\mathcal{A}}^{m}_{\mu}  \right)
\xi_{m} \right|^2 + \widetilde{s} |\xi_m |^2 \right]  +
\sum_{m,n=0}^{2q-1} K_{mn} \left( \epsilon_{\mu\nu\lambda}
\partial_{\nu} \widetilde{\mathcal{A}}^{m}_{\lambda} \right)  \left( \epsilon_{\mu\rho\sigma}
\partial_{\rho} \widetilde{\mathcal{A}}^{n}_{\sigma} \right)  \nonumber \\ &-& i
\overline{\Psi} \gamma_\mu \left(
\partial_\mu + \frac{i}{2} \sum_{m=0}^{q-1}
\widetilde{\mathcal{A}}^m_\mu - \frac{i}{2} \sum_{m=q}^{2q-1}
\widetilde{\mathcal{A}}^m_\mu \right) \Psi, \label{l4}
\end{eqnarray}
where the gauge fields still obeys the constraint
\begin{equation}
\sum_{m=0}^{2q-1} \widetilde{\mathcal{A}}^m_\mu = 0. \label{lc4}
\end{equation}

\paragraph*{Doped VBS, $q \neq 2~\mbox{(mod 4)}$:} For the
doped VBS case, we have already discussed a situation with the
absence of gapless fermions at the quantum critical point in
Section~\ref{sec:sinf}. However, in Sections~\ref{sec:vbsferm}
and~\ref{sec:undual}, we presented plausible conditions under which
the gapless nodal fermions could survive at the quantum critical
point in this case too. With $q'$ defined as in
Eq.~(\ref{defqprime}), it was required that $q'$ be even, and the
critical theory was the same as $\mathcal{L}_3$ or $\mathcal{L}_4$,
but with $q$ replaced by $q'/2$. A more detailed direct lattice
study of the PSG of fractionalization would be useful to firmly
establish this scenario.

\paragraph*{Doped sF, even $q$:}
For the doped sF case, $q=0~\mbox{(mod 4)}$, was found not to
exhibit a superconductor-insulator transition in
Section~\ref{sec:sfq4}; rather a superconductor-supersolid
transition obtains, which will be noted in the following subsection.
The case $q=2~\mbox{(mod 4)}$ is likely to have monopole terms, and
so we are not able to reach any firm conclusions for this case.

\subsubsection{Superconductor-supersolid transition}
\label{sec:ss}

All the cases considered in this paper allow a
superconductor-supersolid transition: this can happen if a
vortex-anti-vortex pair condensate appears before the condensation
of single vortices. This pair condensate transforms just like a
conventional Landau density-wave or VBS order parameter, and the
critical theory can be developed in a traditional order parameter
framework. Such critical theories for the superconductor-supersolid
transition were discussed at some length in Refs.~\cite{bfn}
and~\cite{vs}.

\paragraph*{Doped sF, $q=0~\mbox{(mod 4)}$:}
One of our surprising results was that a superconductor-supersolid
transition was not merely optional for a particular case, but
required as the first transition out of the translationally
invariant superconductor (this conclusion is based upon an
assumption on neglect of monopoles which has not been firmly
established). This case was for the doped sF state with
$q=0~\mbox{(mod 4)}$. We showed in Section~\ref{sec:sfq4} that the
`non-relativistic' nature of the vortex action in this case promoted
the formation of vortex-anti-vortex bound states which would
condense first. The PSG of the vortex theory also placed some
unusual constraints on the nature of the ordering in the supersolid:
unless some higher order couplings where anomalously large, it was
found that the supersolid in the doped sF case could have density
modulations only at wavevectors $(2 \pi p/q)(m,n)$ with $m+n$ odd.

\subsection{Discussion}

To conclude, we have found a remarkable richness in the low energy
spectra and quantum phase transitions of a ``conventional''
two-dimensional $d$-wave superconductor. The vortices of the
superconducting state, when considered as bona fide,
quantum-mechanical, quasiparticle excitations, can have surprising
variety of wavefunctions on the lattice. These wavefunctions
encapsulate the structure of conventional density-wave or VBS orders
in proximate phases. Combining the quantum mechanics of these vortex
quasiparticles with the fermionic Bogoliubov quasiparticles is a
problem of considerable complexity: this paper has described such
theories in two of the simplest cases, and the results are
summarized above.

We conclude by briefly discussing connections to experiments. The
many experimental observations
\cite{hoffman,fischer,fang,ali,mcelroy,hanaguri,steiner,tranquada,abbamonte,hashimoto}
of periodic modulations in the LDOS or spin excitation spectra
clearly call for attention to such competing orders: we have shown
here, following Ref.~\cite{psgbosons}, how such weak modulations
appear naturally as an inevitable, but ancillary, consequence of
quantum fluctuations of the vortices. Recent observations
\cite{valla} of the electronic spectrum of such modulated states
show evidence for gapless nodal fermions: this possibly points to
proximity to deconfined critical points with such fermions discussed
above. For the doped sF case, we also found an unexpected selection
of wavevectors of the modulations, which was just noted above: for
$\delta=1/8$, the density modulations were dominant at wavevectors
$(\pi/8)(m,n)$ with $m+n$ odd. This feature of the sF case appears
to be in conflict with existing observations
\cite{hanaguri,abbamonte}.

\section{Acknowledgements}

This research was supported by the National Science Foundation under
grants DMR-04-57440 (L.B.), DMR-0537077 (S.S.), and the Packard
Foundation (L.B.).

\end{document}